\documentclass[aps,nofootinbib,superscriptaddress,twocolumn,amsmath,amssymb]{revtex4-1}
\usepackage{graphicx}
\usepackage{hyperref}
\usepackage{multirow}
\usepackage{subcaption}
\usepackage{cancel}
\usepackage[force]{feynmp-auto}
\usepackage{natbib}
\usepackage{xcolor}

\newcommand{\half}{\tfrac{1}{2}}
\newcommand{\tenth}{\tfrac{1}{10}}

\newcommand{\pti}[1]{p_{\textrm{T},#1}}
\newcommand{\ecf}[1]{\textrm{ECF}_{#1}}
\newcommand{\kb}[2]{\left(\kappa=#1, \beta=#2 \right)}
\newcommand{\rinv}{r_{\textrm{inv}}}

\newcommand{\midrule}{\colrule}
\newcommand{\bottomrule}{\botrule}
\newcommand{\cmidrule}[1]{\cline{#1}}

\newcommand{\gPlot}[4]{\begin{gathered}\includegraphics[width=#1\textwidth]{fig_#2_#3_#4.pdf}\end{gathered}}
\newcommand{\pmkb}[2]{\begin{pmatrix} \kappa=#1\\ \beta=#2 \end{pmatrix}}
\newcommand{\gEq}[6]{\left(\gPlot{#1}{#2}{#3}{#4}\right)^{\footnotesize{\pmkb{#5}{#6}}}}

\renewcommand{\phi}{\varphi}

\newcommand{\rref}[1]{Ref. \cite{#1}}
\newcommand{\rrefs}[1]{Refs. \cite{#1}}
\newcommand{\Fig}[1]{Fig.~\ref{#1}}

\newcommand{\Tab}[1]{Table~\ref{#1}}

\newcommand{\App}[1]{App.~\ref{#1}}

\newcommand{\Eq}[1]{Eq.~\eqref{#1}}

\newcommand{\sch}[1]{ #1}
\newcommand{\rev}[1]{{#1}}
\newcommand{\rvv}[1]{ #1}

\newcommand{\gev}{\text{GeV}}
\date{\today}

\begin{document}
\preprint{}

\title{Learning to Identify Semi-Visible Jets}

\begin{abstract}
	We train a network to identify jets with fractional dark decay (semi-visible jets) using the pattern of their low-level jet constituents, and explore the nature of the information used by the network by mapping it to a space of jet substructure observables. Semi-visible jets arise from dark matter particles which decay into a mixture of dark sector (invisible) and Standard Model (visible) particles.  Such objects are challenging to identify due to the complex nature of jets and the alignment of the momentum imbalance from the dark particles with the jet axis, but such jets do not yet benefit from the construction of dedicated theoretically-motivated jet substructure observables. A deep network operating on jet constituents is used as a probe of the available information and indicates that  classification power not captured by current high-level observables arises primarily from low-$p_\textrm{T}$ jet constituents.
\end{abstract}

\author{Taylor Faucett}
\affiliation{Department of Physics and Astronomy, University of California, Irvine CA}
\author{Shih-Chieh Hsu}
\affiliation{Department of Physics, University of Washington, Seattle WA}
\author{Daniel Whiteson}
\affiliation{Department of Physics and Astronomy, University of California, Irvine CA}

\maketitle

\section{Introduction}
\label{sec:intro}

The microscopic nature of dark matter (DM) remains one of the most pressing open questions in modern physics~\cite{RevModPhys.90.045002, Tait2018}, and a robust program of experimental searches for evidence of its interaction with the visible sector~\cite{PhysRevLett.108.211804, atlas_dm, CMS_dm, PhysRevLett.110.011802, PhysRevLett.108.261803, Tait2013, PhysRevD.77.115020, PhysRevD.95.034001, Aaij:2017rft} described by the Standard Model (SM). These experiments typically assume that DM is neutral, stable and couples weakly to SM particles; in collider settings this predicts detector signatures in which weakly-produced DM particles are invisible, evidenced only by the imbalance of momentum transverse to the beam. No evidence of DM interactions has been observed to date.

However, while these assumptions are reasonable, the lack of observation motivates the exploration of scenarios in which one or more of them are relaxed.  Specifically, if DM contains complex strongly-coupled hidden sectors, \rvv{such as appear in many Hidden-Valley models~\cite{Strassler:2006im,Han:2007ae}}, it may lead to the production of stable or meta-stable dark particles within hadronic jets~\cite{Cohen2015, Cohen2017, Kar:2020bws, Bernreuther:2020vhm}. Depending on the portion of the jet which results in dark-sector hadrons, it may be only {\it semi-visible} to detectors, leading to a unique pattern of energy deposits, or jet substructure.

A robust literature exists for the identification of jet substructure~\cite{Larkoski2014SoftDrop, Larkoski:2013uj,Kogler2019,Lu:2022cxg}, with applications to boosted $W$-boson~\cite{Baldi:2016fql,PhysRevD.101.053001,Aaboud:2019uz}, Higgs boson~\cite{Chung2021} and top-quark tagging~\cite{Aaboud:2019uz}. \rev{Typically, observables are designed to discriminate between the energy deposition patterns left by jets which result from a single hard parton and the patterns left by  jets which result from several collimated hard partons, as can be produced from the decay of a massive boosted particle}. While these observables have some power when adapted to the task of identifying semi-visible jets~\cite{Kar:2020bws,Cohen:2020afv}, no observables have yet been specifically designed to be sensitive to the unique energy patterns of semi-visible nature of jets.

In parallel, the rapid development of machine learning to the analysis of jet energy depositions~\cite{Larkoski:2013uj,Komiske:2017aww,Baldi:2016fql} demonstrated that jet tagging strategies, including those for semi-visible jets, can be learned directly from lower-level jet constituents without the need to form physics-motivated high level observables\rev{~\cite{Bernreuther:2020vhm, Dillon:2022mkq}}.  Such learned models are naturally challenging to interpret, validate or quantify uncertainties, especially given the high-dimensional nature of their inputs. \rvv{In the case of semi-visible jets, extra care must be taken when drawing conclusions from low-level details, many of which may depend on specific theoretical parameter choices as well as the approximations made during modeling of hadronization~\cite{Cohen:2020afv}}.  However, techniques have been recently demonstrated~\cite{Faucett2020,Collado2021,Collado2021b,Bradshaw:2022qev} to translate the learned model into interpretable high-level observables, \rvv{which can provide guidance regarding the nature and robustness of the information being used for classification.}

In this paper, we present a  study of machine learning models trained to distinguish semi-visible jets from QCD background jets using the patterns of their low-level jet constituents. We compare the performance of models which use low-level constituents to those which use the set of existing high-level observables to explore where the existing HL observables do and do not capture all of the relevant information. Where gaps exist, we attempt to translate~\cite{Faucett2020} low-level networks into networks which use a small set of interpretable observables which replicate their decisions and performance.  Interpretation of these observables can yield insight into the nature of the energy deposition inside semi-visible jets.

\section{Semi-visible jets}
\label{sec:theory}
\begin{figure*}
	\centering
	\begin{subfigure}[t]{0.45\textwidth}
		\centering
		\includegraphics[width=0.95\textwidth]{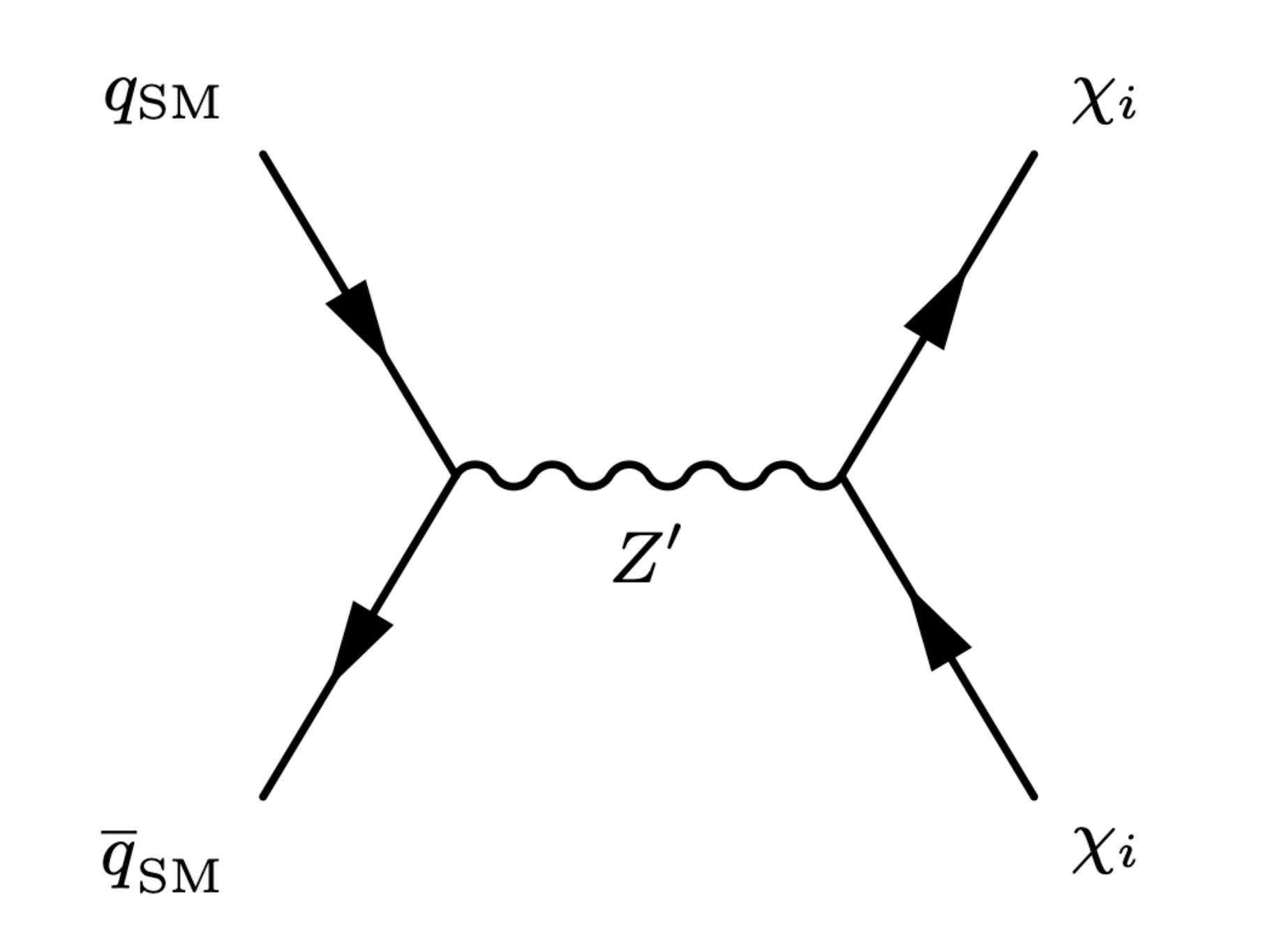}
		\caption{s-channel}
		\label{fig:feyn_s}
	\end{subfigure}\vspace{3em}
	\begin{subfigure}[t]{0.45\textwidth}
		\centering
			\includegraphics[width=0.95\textwidth]{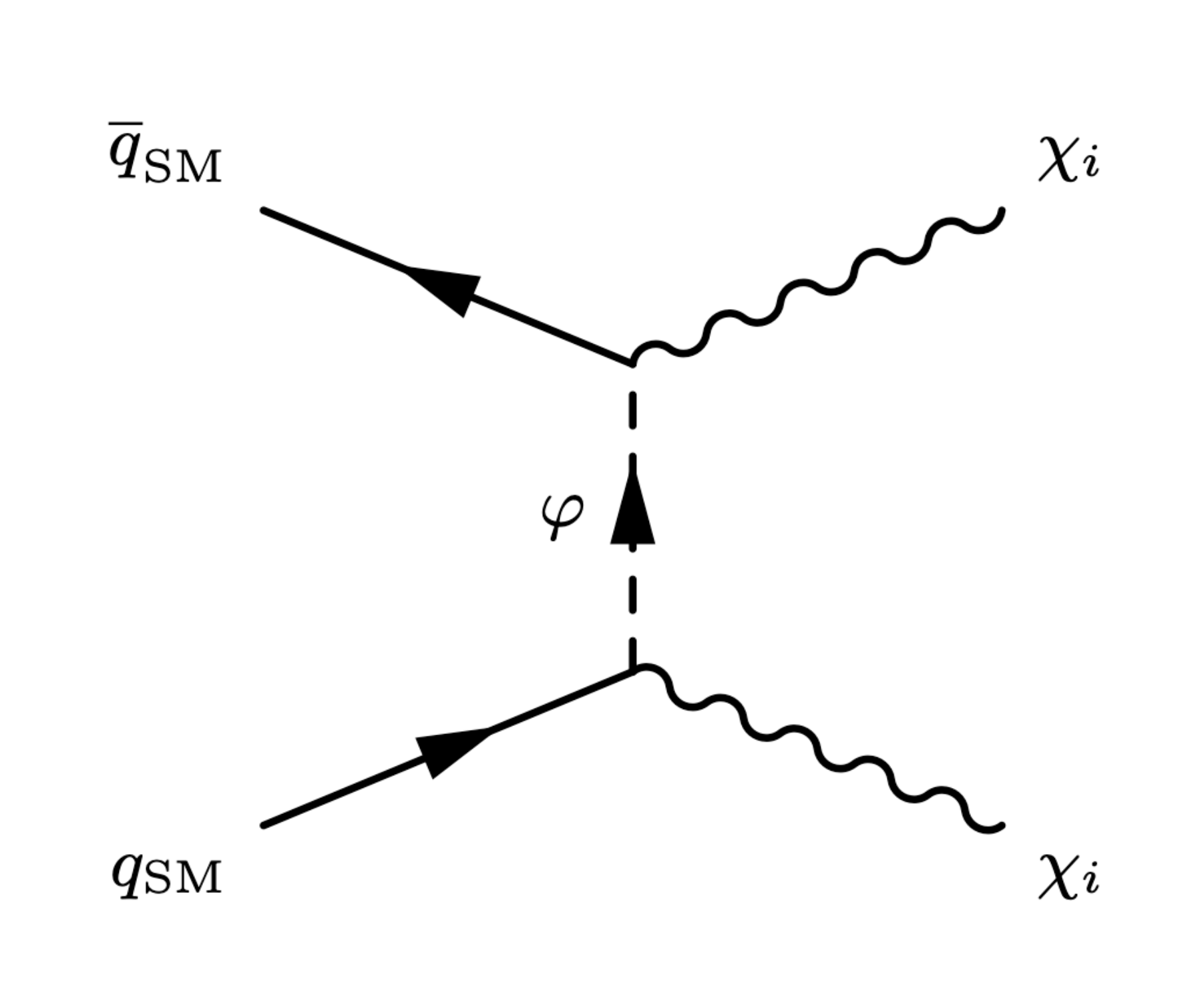}
		\caption{t-channel}
		\label{fig:feyn_t}
	\end{subfigure}
	\caption{ Feynman diagrams for $s$-channel and $t$-channel  pair-production of dark-sector quarks $\chi_i$ which lead to semi-visible jet production.}
\end{figure*}

Following \rrefs{Cohen2015,Kar:2020bws}, we consider \rev{pair} production of dark-sector quarks of several flavors $\left(\chi_i = \chi_{1,2}\right)$ via a messenger sector which features a $Z^{\prime}$ gauge boson in an $s$-channel process (\Fig{fig:feyn_s}) or scalar mediator $\phi$ in the $t$-channel process (\Fig{fig:feyn_t}) that couples to both SM and DM sectors \rev{and leads to a dijet signature}. The dark quarks produce QCD-like dark showers, involving many dark quarks and gluons which produce dark hadrons, some of which are stable or meta-stable and some of which decay into SM hadrons via an off-shell process.

\begin{figure}
	\centering
	\begin{subfigure}[t]{0.22\textwidth}
		\centering
		\includegraphics[width=\textwidth]{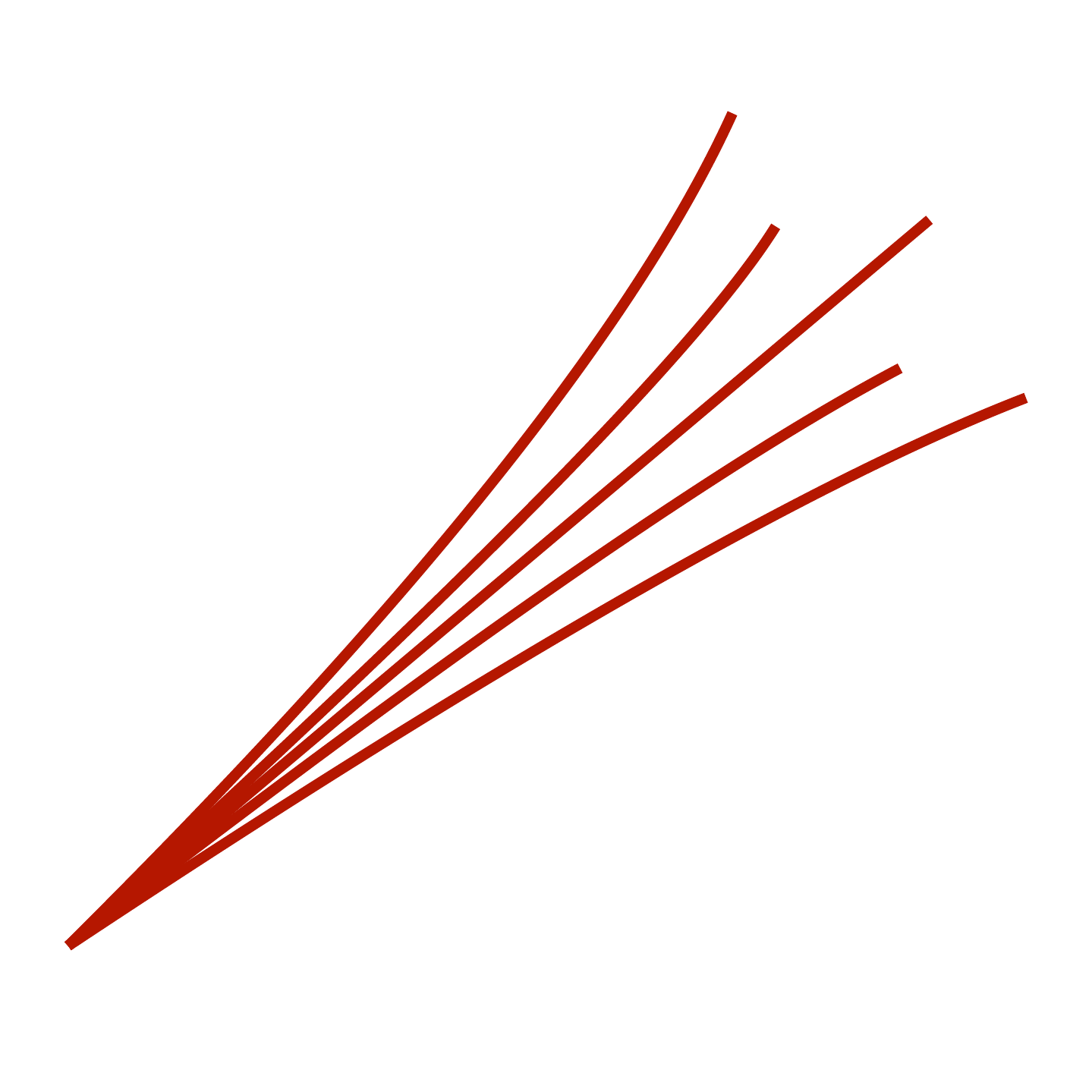}
		\caption{Rapid Decay (Visible), $\rinv=0$}
		\label{fig:RapidDecay}
	\end{subfigure}
	\begin{subfigure}[t]{0.22\textwidth}
		\centering
		\includegraphics[width=\textwidth]{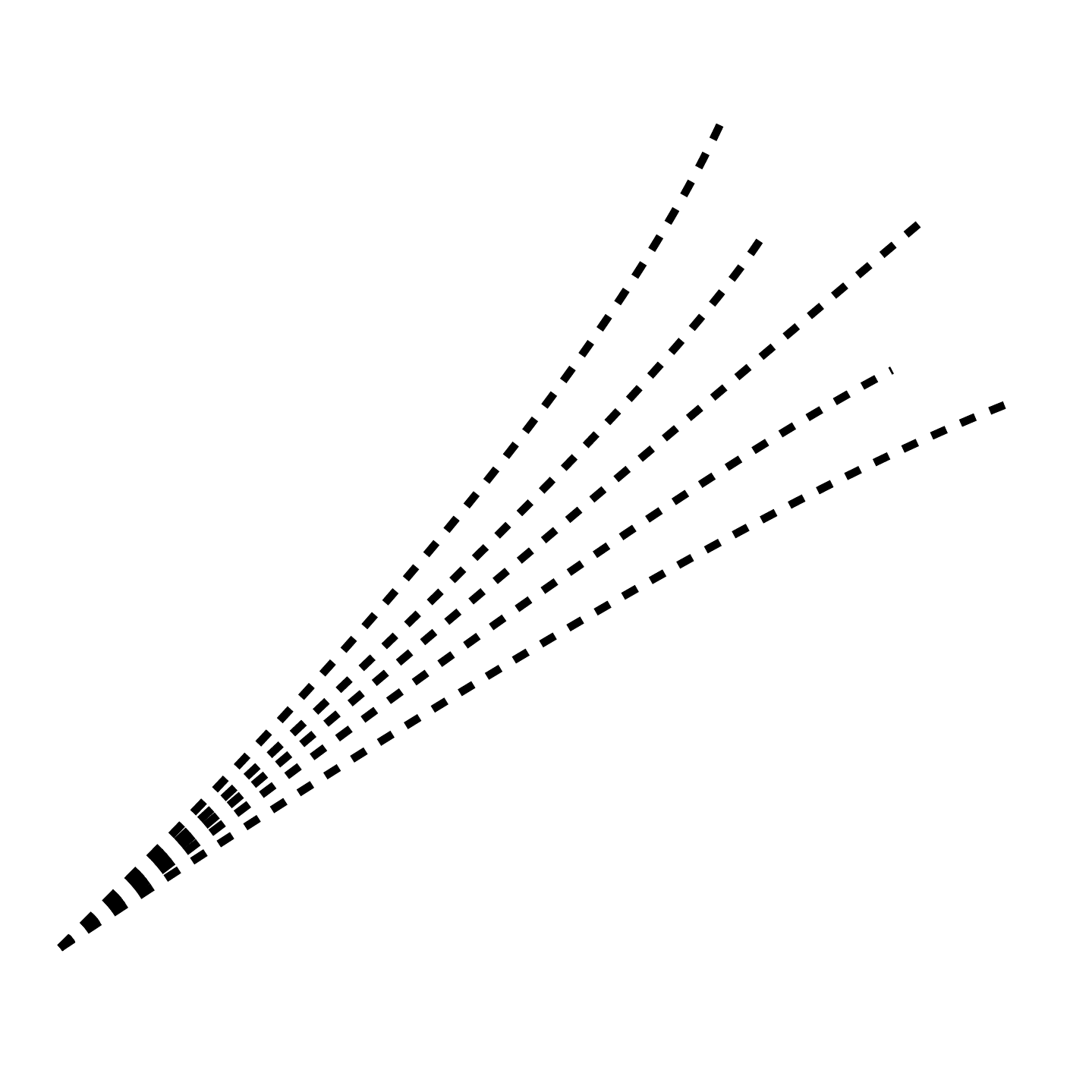}
		\caption{Stable Dark Jet (Invisible), $\rinv=1$}
		\label{fig:StableDark}
	\end{subfigure}
	\begin{subfigure}[t]{0.3\textwidth}
		\centering
		\includegraphics[width=\textwidth]{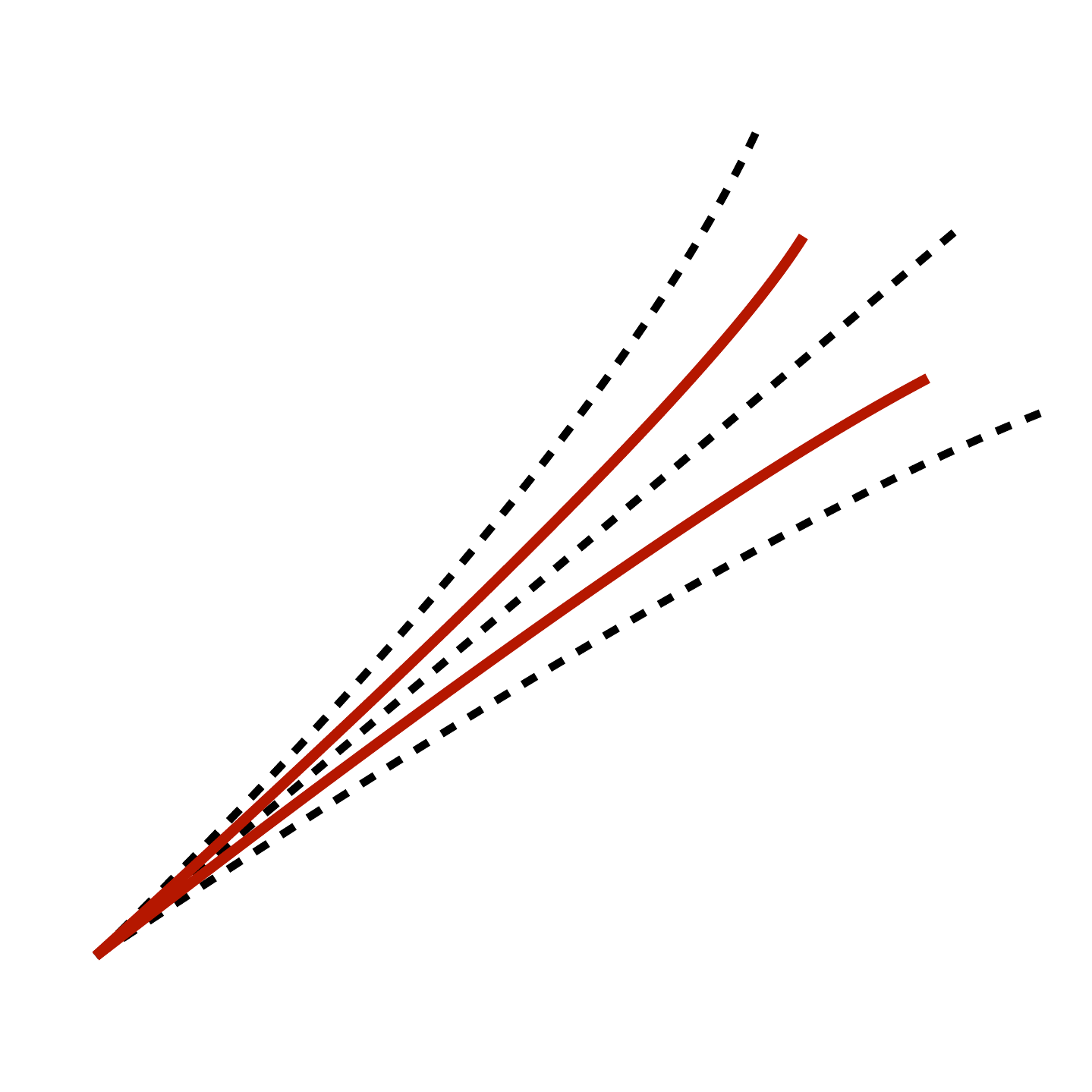}
		\caption{Fractional Decay (Semi-Visible), $\rinv \in (0,1)$}
		\label{fig:SVJ}
	\end{subfigure}
	
	\caption{ Depictions of jets produced with varying visible (SM, solid red) and invisible (dark sector, dashed black) components.}
	\label{fig:dark-modes}
\end{figure}

The detector signature of the resulting semi-visible jet (SVJ) depends on the lifetime and stability of the dark hadrons, leading to several possibilities (see
\Fig{fig:dark-modes}). Though the physics is complex and \rev{depends} on the details of the dark sector structure, a description of the dark and SM hadrons produced by a DM model quark can be encapsulated in the quantity $\rinv$, the ratio of dark stable hadrons to all hadrons in the jet:

\begin{equation}
	\rinv \equiv \left\langle \frac{\textrm{\# of stable dark hadrons}}{\textrm{\# of  hadrons}} \right\rangle
	\label{eq:rinv}
\end{equation}

An invisible fraction of $\rinv=0$ corresponds to a dark quark which produces a jet consisting of only visible hadrons, as depicted in the \emph{Rapid Decay} example given in \Fig{fig:RapidDecay}. Alternatively, an invisible fraction of $\rinv=1$ describes a stable dark jet (\Fig{fig:StableDark}), in which the dark quark hadronizes exclusively in the dark sector.  For any intermediate value of $\rinv$, jets contain a visible and invisible fraction, leading to $\cancel{E}_{\textrm{T}}$ along the jet axis(\Fig{fig:SVJ}).

\section{Sample Generation and Data Processing}
\label{sec:generation}
Samples of simulated events with semi-visible jets are generated using the modified Hidden Valley~\cite{Strassler2007} model described in ~\cite{Cohen2015,Cohen2017} for both an $s$-channel (\Fig{fig:feyn_s}) and $t$-channel (\Fig{fig:feyn_t}) process.  Samples of simulated  $pp\to Z^{\prime} \to \chi_1 \overline{\chi}_1$ and $pp\to \phi \to \chi_1 \overline{\chi}_1$ events are produced in  proton-proton collisions at a center-of-mass energy of $\sqrt{s} = 13\, \text{TeV}$ in MadGraph5~\cite{Alwall2011} (v2.6.7) with xqcut=100 and the NNPDF2.3 LO PDF set~\cite{Skands2014}. The mediator mass is set to $1.5\, \textrm{TeV}$ and the dark quark mass to $M_{\chi_1}=10\, \textrm{GeV}$.  Up to two extra hard jets due to radiation are generated and MLM-matched~\cite{Mangano2002MultijetStudy}, to facilitate  comparison with existing studies. 
\sch{Invisible fraction, showering and hadronization are performed with Pythia8 v8.244~\cite{pythia}. In particular, the following parameters are set: the dark confinement scale $\Lambda_d = 5\, \textrm{GeV}$; the number of dark colors $N_c = 2$; the number of dark flavors $N_f = 1$; and the intermediate dark meson $\rho_d$ mass of $m_{\rho_d} = 20\, \textrm{GeV}$ and the dark matter $\pi_d$ mass of $m_d = 9.9\, \textrm{GeV}$. Distinct sets were generated for invisible fractions of $\rinv\in[0.0, 0.3, 0.6]$ via configurations of the branching ratio of decay process $\rho_d\to \pi_d \pi_d$. }
Detector simulation and reconstruction are conducted in Delphes v3.4.2~\cite{deFavereau:2013fsa} using the default ATLAS card. A sample of SM jets from a typical QCD process is generated from the $p p \to j j$ process. The same simulation chain as described above is applied to the SM jets. 

Jets are built from calorimeter energy deposits and undergo the jet trimming procedure described in \rref{Krohn2010JetTrimming} with the anti-$k_\textrm{T}$~\cite{Cacciari2008} clustering algorithm from pyjet~\cite{Cacciari2012}, using a primary jet-radius parameter of $R=1.0$ and subjet clustering radius of $R_\text{sub} = 0.2$ and $f_{\text{cut}}=0.05$. \rev{The threshold on $f_{\text{cut}}$ effectively removes constituents in subjets with $p_\textrm{T}$ that is less than 5\% of the jet $p_\textrm{T}$}. Leading jets are required to have $p_{\textrm{T}} \in [300, 400]\, \gev$. For each event generated, the leading jet is selected and truth-matched to guarantee the presence of a dark quark within the region of $\Delta R < 1$.

After all selection requirements, $2\times 10^{6}$ simulated jets remain with a 50/50 split between SVJ signal and QCD background. To avoid inducing biases from artifacts of the generation process, signal and background events are weighted such that the distributions in $p_{\textrm{T}}$ and $\eta$ are uniform; see ~\Fig{fig:reweight}.

\subsection{High-level Observables}
\label{sec:hl}
A large set of jet substructure observables~\cite{Kogler2019, Marzani2019, Larkoski2017} have been proposed for tasks different from the focus of this study, that of identifying jets with multiple hard subjets.  Nevertheless, these observables may summarize the information content in the jet in a way that is relevant for the task of identifying semi-visible jets~\cite{Kar:2020bws}, and so serve as a launching point for the search for new observables.

Our set of \rev{high-level (HL)} observables includes 15 candidates: jet $p_{\textrm{T}}$, the Generalized Angularities~\cite{Larkoski:2014vc} of $p_{\textrm{T}}^D$ and Les Houches Angularity (LHA), N-subjettiness ratios $\tau_{21}^{\beta=1}$ and $\tau_{32}^{\beta=1}$~\cite{Thaler2011IdentifyingN-subjettinessb}, and Energy Correlation function ratios\cite{Larkoski:2013uj} $C_2^{\beta=1}$, $C_2^{\beta=2}$, $D_2^{\beta=1}$, $D_2^{\beta=2}$, $e_2$, $e_3$, as well as jet $\textrm{width}$, jet $e_\textrm{mass}$, constituent multiplicity and the splitting function~\cite{Larkoski2014SoftDrop} $z_g$.  In each case, observables are calculated from trimmed jet constituents described above.  Definitions and distributions for each high-level observable are provided in \App{app:jss_hist} with re-weighting applied as described above.

\section{Machine Learning and Evaluation}
\label{sec:ml_performance}
For both the low-level (LL) trimmed jet constituents and high-level jet substructure observables, a variety of networks and architectures are tested. 

Due to their strong record in previous similar applications~\cite{Faucett2020, Collado2021, Collado2021b,Bernreuther:2020vhm}, a deep neural network using dense layers is trained on the high-level observables. We additionally consider XGBoost\cite{Chen:2016:XST:2939672.2939785}, which has shown strong performance in training high-level classifiers with jet substructure~\cite{Bourilkov2019, cornell2021boosted}, as well as LightGBM\cite{NIPS2017_6449f44a} which has demonstrated power in class separation on high-level features. LightGBM has the strongest classification performance among these networks which use high-level features; see Tab. \ref{tab:hl_performance_results}.

\begin{table}
	\caption{Summary of performance (AUC) in the SVJ classification task for several networks using high-level features, for the six simulated scenarios, three choices of invisible fraction $\rinv$ for both the $s$-channel and $t$-channel processes. 	Statistical uncertainty in each case is less than $\pm 0.002$ with a 95\% confidence, measured using bootstrapping over 200 models.}
	\label{tab:hl_performance_results}
	\centering
	\begin{tabular}{lccc}\toprule
        $s$-channel \hspace{3em} & \multicolumn{3}{c}{AUC}\\ \cmidrule{2-4}
        Model & $\rinv=0.0$ & 0.3 & 0.6\\ \midrule
		LightGBM  & 0.861  &  0.803 & 0.736 \\
		XGBoost   & 0.861  & 0.803 &  0.736 \\
		DNN  & 0.860  & 0.799  & 0.734 \\
		\bottomrule
        $t$-channel& \multicolumn{3}{c}{AUC}\\ \cmidrule{2-4}
        Model & $\rinv=0.0$ & 0.3 & 0.6\\ \midrule
        LightGBM  & 0.808 & 0.755 & 0.683 \\
        XGBoost  & 0.806  & 0.755 & 0.681 \\
        DNN & 0.801 & 0.726 & 0.656 \\
		\bottomrule
	\end{tabular}
\end{table}

\begin{table}
	\caption{Summary of performance (AUC) in the SVJ classification task for several networks using low-level constituents, for the six simulated scenarios, three choices of invisible fraction $\rinv$ for both the $s$-channel and $t$-channel processes. 	Statistical uncertainty in each case is less than $\pm 0.002$ with a 95\% confidence, measured using bootstrapping over 200 models.}
	\label{tab:ll_performance_results}
	\centering
	\begin{tabular}{lccc}\toprule
        $s$-channel \hspace{3em} & \multicolumn{3}{c}{AUC}\\ \cmidrule{2-4}
        Model & $\rinv=0.0$ & 0.3 & 0.6\\ \midrule
        PFN & 0.866 & 0.822 & 0.776\\
		EFN & 0.849 & 0.795 & 0.735\\
		CNN & 0.855 & 0.792 & 0.740\\
		\bottomrule
        $t$-channel& \multicolumn{3}{c}{AUC}\\ \cmidrule{2-4}
        Model & $\rinv=0.0$ & 0.3 & 0.6\\ \midrule
        PFN & 0.806 & 0.754 & 0.697\\
		EFN & 0.796 & 0.741 & 0.672\\
		CNN & 0.791 & 0.739 & 0.663\\
		\bottomrule
	\end{tabular}
\end{table}

In the case of classifiers which use low-level constituents, convolutional neural networks on jet images are considered~\cite{Faucett2020, Collado2021, Collado2021b, Baldi:2016fql, Cogan:2014oua, deOliveira:2015xxd}. Given the specific task of classifying jet substructure observables and their use for a similar task in \rref{Collado2021b}, Energy Flow Networks (EFN) and Particle Flow Networks (PFN) are also applied~\cite{Komiske:2018cqr}, and found to significantly out-perform convolutional networks, with the PFN emerging as the most powerful network; see \Tab{tab:ll_performance_results}. 

Receiver operating characteristic (ROC) curves for both the PFN and LightGBM high-level models are given in \Fig{fig:roc_performance}. Additional details for the network training and hyperparameter selections are provided in \App{sec:ml_arch}.

\subsection{Performance Comparison}
\label{sec:ml-results}

\begin{figure}
	\centering
	\includegraphics[width=0.48\textwidth]{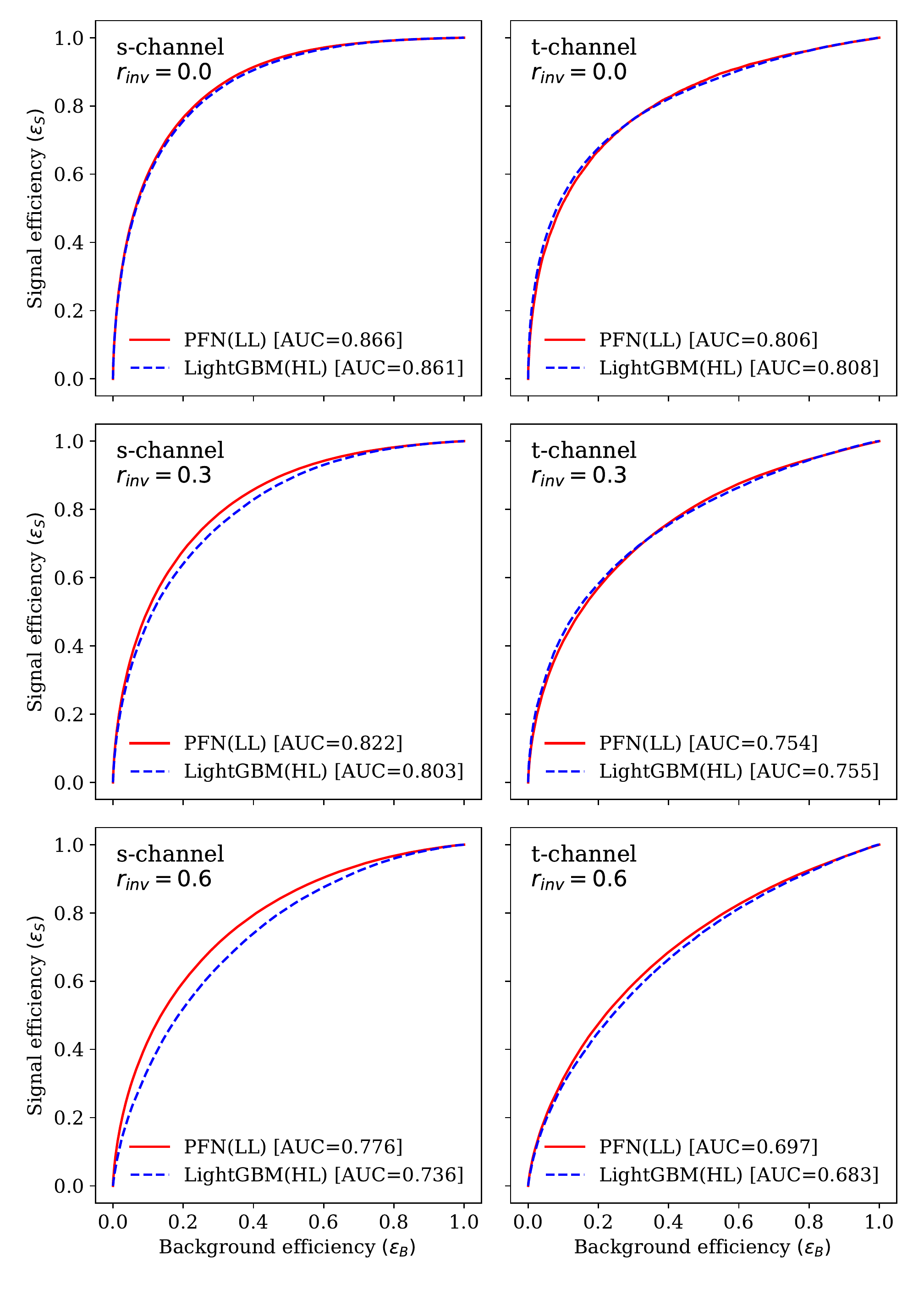}
	\caption{ Comparison of performance between a network which uses low-level constituents (PFN in solid red) and one which uses high-level jet observables (LightGBM, dashed blue). Shown are the background rejection (inverse of background efficiency) versus signal efficiency for the six simulated cases: $s$-channel and $t$-channel production for three values of $r_\textrm{inv}.$ Statistical uncertainty on AUC in each case is less than $\pm 0.002$ at 95\% confidence level, measured using bootstrapping over 200 models.}
	\label{fig:roc_performance}
\end{figure}

We compare the SVJ classification performance of the most powerful networks based on high-level or low-level input features, through calculations of the Area under the ROC Curve (AUC); see Fig~\ref{fig:roc_performance} and \Tab{tab:performance_results}.

Note that the \rev{high-level (HL)} observables are calculated directly from the low-level constituents with no additional information. If HL features extract all of the relevant information for the classification task, networks which use them as inputs should be able to match the performance of the networks which use the LL information directly, which we take as a probe of the power of the available information. \rev{If networks using LL information surpass the performance of those using HL information, it suggests that information exists in the LL constituents which is not being captured by the HL observables}. One might consider directly applying networks based on LL information to the classification task~\cite{Bernreuther:2020vhm}, but this presents challenges in calibrating, validating and quantifying uncertainties on their high-dimensional inputs.  Instead, a performance gap suggests the possibility that an additional HL observable might be crafted to take advantage of the overlooked information. 

In each of the six cases explored here, the networks which use low-level information match or exceed the performance of networks which use high-level jet observables. Significant performance gaps are seen in the $\rinv=0.6$ case, where the AUC between the LL and HL networks is 0.040 and 0.014 for the $s$-channel and $t$-channel cases, as well as in the and $\rinv=0.3$ invisible fraction in the $s$-channel process, where the gap is 0.019.  In the $\rinv=0.0$ $s$-channel scenario, a small gap is seen, though larger than the statistical uncertainty. In the other two cases, the LL and HL networks achieve essentially equivalent performance, strongly suggesting that the HL features have captured the relevant information in the LL constituents. \rev{As these observables were not designed for this task, it was not a priori clear that they would summarize all of the available and relevant information.}

However, one can also assess the difference between the LL and HL networks using other metrics than AUC. The Average Decision Ordering (ADO) metric, see Eq.~\ref{eq:DO_def}, measures the fraction of input pairs in which two networks give the same output ordering. Even in cases where the AUC is equivalent, the ADO between the LightGBM and PFN (\Tab{tab:performance_results}) is well below 1, suggesting that while their classification performance is the same, they arrive at it using distinct strategies. This indicates that there may be room to improve the classification accuracy by considering a network which uses both sets of features.
\begin{table*}
    \centering
	\caption{Summary of performance (AUC and ADO) in the SVJ classification task for various network architectures and input features. Statistical uncertainty in each case is less than $\pm 0.002$ at 95\% confidence level, measured using bootstrapping over 200 models.}
	\label{tab:performance_results}
	\begin{tabular}{lcccccc}\toprule
	    \multicolumn{7}{c}{$s$-channel}\\
		\multirow{2}{*}{Model[Features]}           & \multicolumn{2}{c}{$\rinv=0.0$} & \multicolumn{2}{c}{$\rinv=0.3$} & \multicolumn{2}{c}{$\rinv=0.6$}                              \\
		\cmidrule{2-3} \cmidrule{4-5} \cmidrule{6-7} & ADO[$\cdot$,PFN]                      & AUC                             & ADO[$\cdot$,PFN]                      & AUC   & ADO[$\cdot$,PFN] & AUC   \\ \midrule
		PFN[LL]                                     & 1                               & 0.866                           & 1                               & 0.822 & 1          & 0.776 \\
		LightGBM[HL]                                     & 0.858                           & 0.861                           & 0.839                           & 0.803 & 0.818      & 0.736 \\
		\hline
		LL-HL gap       & & \bf{0.005} & & \bf{0.019} & & \bf{0.040}\\
		\bottomrule
		\multicolumn{7}{c}{\vspace{0.5em}}\\ \toprule
		\multicolumn{7}{c}{$t$-channel}\\
		\multirow{2}{*}{Model[Features]}           & \multicolumn{2}{c}{$\rinv=0.0$} & \multicolumn{2}{c}{$\rinv=0.3$} & \multicolumn{2}{c}{$\rinv=0.6$}                              \\
		\cmidrule{2-3} \cmidrule{4-5} \cmidrule{6-7} & ADO[$\cdot$,PFN]                      & AUC                             & ADO[$\cdot$,PFN]                      & AUC   & ADO[$\cdot$,PFN] & AUC   \\ \midrule
		PFN[LL]                                      & 1                               & 0.806                           & 1                               & 0.754 & 1          & 0.697 \\
		LightGBM[HL]                                     & 0.844                           & 0.808                           & 0.805                           & 0.755 & 0.787      & 0.683 \\
			\hline
		LL-HL gap      & & \bf{-0.002} & & \bf{-0.001} & & \bf{0.014}\\
		\bottomrule
	\end{tabular}
\end{table*}

\section{Finding New Observables}
\label{sec:analysis}
The studies above reveal that models which use low-level constituents as inputs provide the overall best classification performance.  However, our objective is not merely to find the classifier with the optimal statistical performance with difficult-to-assess systematic uncertainties. Rather, we seek to understand the underlying physics used by the PFN and to translate this information into one or more meaningful physical features, allowing us to extract insight into the physical processes involved and assign reasonable systematic uncertainties.  In this section, we search for these additional high-level observables among the Energy Flow Polynomials~\cite{Komiske:2017aww} (EFP), which form a complete basis for jet observables.

\subsection{Search Strategy}
\label{sec:search-strategy}
Interpreting the decision making of a black-box network is a notoriously difficult problem. For the task of jet classification, we apply the guided search technique utilized in past jet-related interpretability studies~\cite{Faucett2020, Collado2021b, Collado2021, Bradshaw:2022qev,Lu:2022cxg}. In this approach, new HL observables are identified among the infinite set of Energy Flow Polynomials (EFPs) which exist as a complete linear basis for jet observables. In the EFP framework, one-dimensional observables are constructed as nested sums over measured jet constituent transverse momenta $\pti{i}$ and scaled by their angular separation $\theta_{ij}$. 

These parametric sums are described as the set of all isomorphic multigraphs where:
\begin{align}
	\text{each node}          & \Rightarrow \sum_{i = 1}^N z_i, \label{eq:EFP_node}          \\
	\text{each $k$-fold edge} & \Rightarrow \left(\theta_{ij}\right)^k \label{eq:EFP_edge} .
\end{align}
and where each graph can be further parameterized by values of $(\kappa, \beta)$,
\begin{align}
	(z_i)^\kappa      & = \left(\frac{\pti{i}}{\sum\limits_j \pti{j}} \right)^\kappa, \label{eq:EFP_z}                              \\
	\theta^\beta_{ij} & = \left(\Delta \eta_{ij}^2 + \Delta \phi_{ij}^2 \right)^{\beta/2}. \label{eq:EFP_theta}
\end{align}
Here, $\pti{i}$ is the transverse momentum of the trimmed jet for constituent $i$, and $\Delta\eta_{ij}$ ($\Delta\phi_{ij}$) is pseudorapidity (azimuth) difference between constituents $i$ and $j$.   As the EFPs are normalized, they capture only the relative information about the energy deposition. For this reason, in each network that includes EFP observables, we include as an additional input the sum of $p_{\textrm{T}}$ over all constituents, to indicate the overall scale of the energy deposition.

For the set of EFPs, infrared and collinear (IRC) safety requires that $\kappa = 1$. To more broadly explore the space,  we consider examples with $\kappa \not= 1$ which generate more exotic observables. For the case of EFPs with IRC-unsafe parameters, we select all prime graphs with dimension $d\leq 5$ and $\kappa$ and $\beta$ values of $\kappa \in \left[-2, -1, 0, \half, 1, 2, 4 \right]$ and $\beta \in \left[\tenth, \half, 1, 2, 4 \right]$. Given the form of \Eq{eq:EFP_node} and \Eq{eq:EFP_edge}, the size and relative sign of inputs chosen for $(\kappa,\beta)$ will provide insights into utility of soft/hard $p_{\textrm{T}}$ effects and narrow/wide angle energy distributions.

In principle, the EFP space is complete and any information accessible through constituent-based observables is contained in some combination of EFPs. However, there is no guarantee that an EFP representation will be compact. On the contrary, a blind search through the space can prove time and resource prohibitive. Rather than a brute force search of an infinite space of observables, we take the guided approach of Ref~\cite{Faucett2020}, which uses the PFN as a black-box guide and iteratively assembles a set of EFPs which provide the closest equivalent decision making in a compact and low-dimensional set of inputs. This is done by isolating the space of information in which the PFN and existing HL features make opposing decisions on the same inputs and isolates the EFP which most closely mimics the PFN in that subspace.

Here, the agreement between networks $f(x)$ and $g(x)$ is evaluated over pairs of inputs $(x,x')$ by comparing their relative classification decisions, expressed mathematically as:
\begin{equation}
	\label{eq:DO_def}
	\textrm{DO}[f,g](x,x') = \Theta \Big( \big( f(x) - f(x') \big) \big( g(x) - g(x') \big) \Big),
\end{equation}
and referred to as \emph{decision ordering} (DO). DO$=0$ corresponds to inverted decisions over all input pairs and DO = 1 corresponds to the same decision ordering.  As prescribed in Ref.~\cite{Faucett2020}, we scan the space of EFPs to find the observable that has the highest average decision ordering (ADO) with the guiding network when averaged over disordered pairs.  The selected EFP is then incorporated into the new model of HL features, HL$_{n+1}$, and the process is repeated until the ADO  plateaus.

\section{Guided Iteration}
\label{sec:guided_iteration}

For each of the four scenarios in which a gap is observed between the AUC performance of the HL LightGBM model and the PFN, a guided search is performed to identify an EFP observable which mimics the decisions of the PFN. 

The search results, shown in \Tab{tab:guided_search_results}, identify in each of the four cases an EFP with $d\leq 3$ which boosts the classification performance of the LightGBM classifier when trained with the original 15 HL features as well as the identified EFP observable. In addition, the decision similarity (ADO) between the new HL model and the PFN is increased.    Scans for a second EFP do not identify additional observables which significantly increase performance or similarity. A guided search was also performed with an identical set of $\kappa$ and $\beta$ parameters for EFPs with dimension $d\leq 5$ but no improvements were seen in the case of higher dimensional graph structure.  While the performance and similarity gaps have been reduced, they have not quite been erased.

\begin{table*}
    \caption{ Results of a guided search for high-level (HL) EFP observables which mimic the decision ordering of the PFN, a network based on low-level constituents. For each of the four scenarios in which a gap is observed between the AUC performance of the HL network and the PFN, an EFP is selected to attempt to increase the average decision ordering (ADO) of the HL network with the PFN. Statistical uncertainty in each case is $\pm 0.002$ with 95\% confidence, measured using bootstrapping over 100 models.}
	\label{tab:guided_search_results}
	\begin{tabular}{cc|c|cccc|c}
		\toprule
		       & & HL network & \multicolumn{4}{c|}{HL+EFP network} & PFN \\
		Process     & $\rinv$ & AUC, ADO[\sc{.,pfn}] & EFP & $\kappa$ & $\beta$ & AUC, ADO[\sc{.,pfn}] & AUC \\ \midrule
		$s$-channel & 0.0  & 0.861,\  0.858 & $\gPlot{0.05}{1}{0}{0}$ & -2 &        & 0.864,\  0.863          &    0.866         \\
		$s$-channel & 0.3  & 0.803,\   0.839 & $\gPlot{0.05}{2}{3}{0}$ & 1 &   $\half$     & 0.807,\  0.840          &    0.822         \\
		$s$-channel & 0.6  & 0.736,\   0.818 & $\gPlot{0.05}{3}{2}{0}$ & -1 &  2      & 0.747,\  0.821          &    0.776         \\
		$t$-channel & 0.6  & 0.683,\   0.787 & $\gPlot{0.05}{3}{3}{1}$ & -2 &  $\tenth$      & 0.690,\  0.792          &    0.697         \\\bottomrule
	\end{tabular}
\end{table*}

\subsection{Analysis of the Guided Search Results}
\begin{figure}
	\centering
	\includegraphics[width=0.48\textwidth]{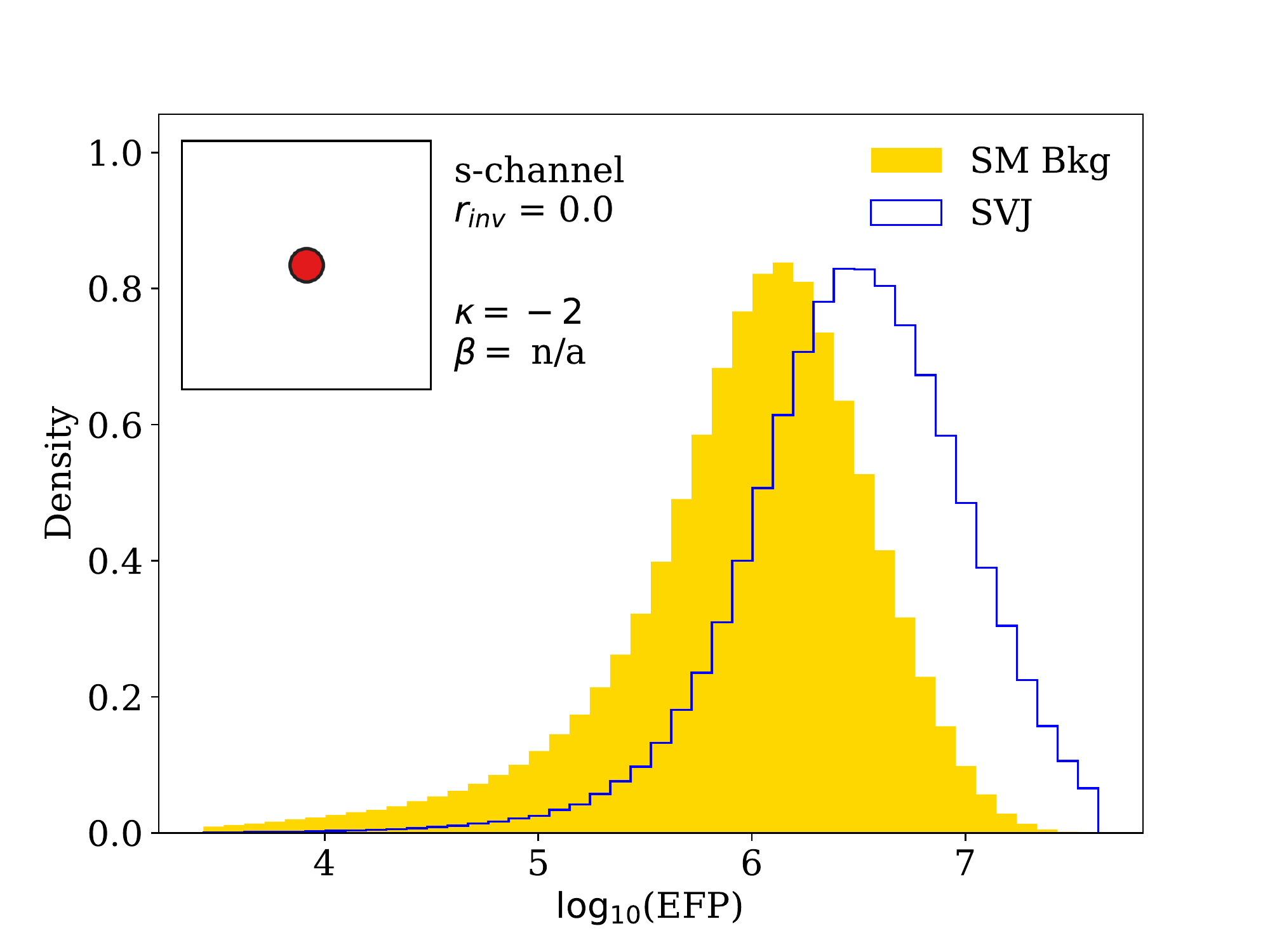}
	\caption{Distribution of the EFP observable selected by the guided search in the $s$-channel $\rinv=0$ scenario, shown for semi-visible jets (SVJ) as well as QCD jets from the Standard Model background (SM Bkg). The inset pane shows the graph corresponding to the EFP. See text for additional details.}
	\label{fig:1_0_0}
\end{figure}

In the $s$-channel process, with invisible fraction \mbox{$\rinv=0.0$}, the addition of a single EFP  closes the small performance gap with the PFN within the statistical uncertainty. The identified EFP in this case is the \emph{dot} graph with the IRC-unsafe energy exponent $\kappa=-2$, expressed as a sum over constituents in \Eq{eq:dot}.
\begin{equation}
	\gEq{0.065}{1}{0}{0}{-2}{\text{n/a}} = \sum\limits_{a=1}^N \frac{1}{z_a^2}
	\label{eq:dot}
\end{equation}
This graph is, in effect, simply a measure of the sum of the inverse $p_{\textrm{T}}^2$ of the jet constituents, and is sensitive to constituents with low $p_{\textrm{T}}$. Distribution of values of this observable for signal and background events are shown in \Fig{fig:1_0_0}, demonstrating good separation between signal and background. 


\begin{figure*}
	\centering
	\begin{subfigure}[t]{0.32\textwidth}
		\centering
		\includegraphics[width=\textwidth]{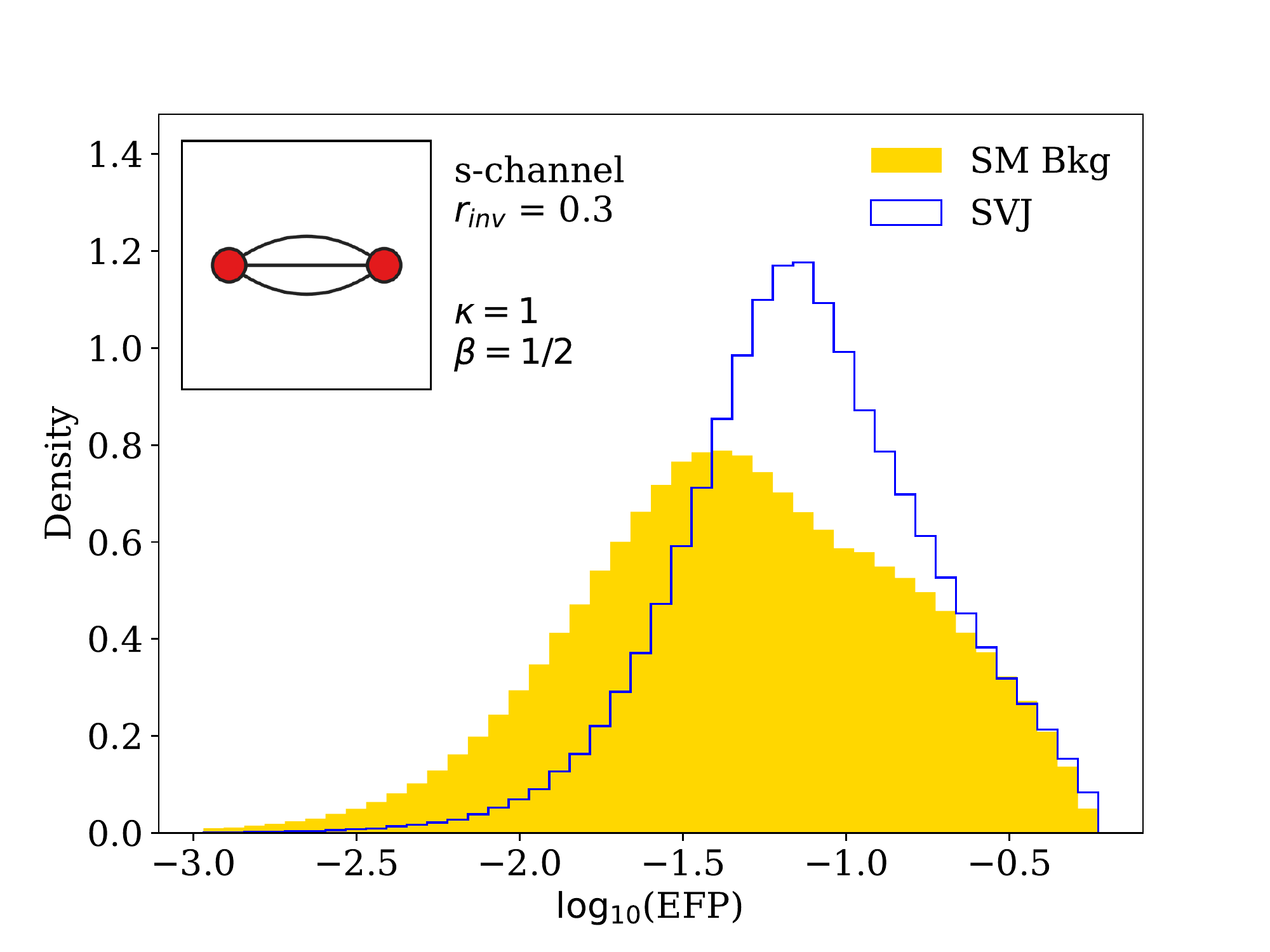}
		\caption{$s$-channel / $\rinv=0.3$}
		\label{fig:graph_s3}
	\end{subfigure}
	\begin{subfigure}[t]{0.32\textwidth}
		\centering
		\includegraphics[width=\textwidth]{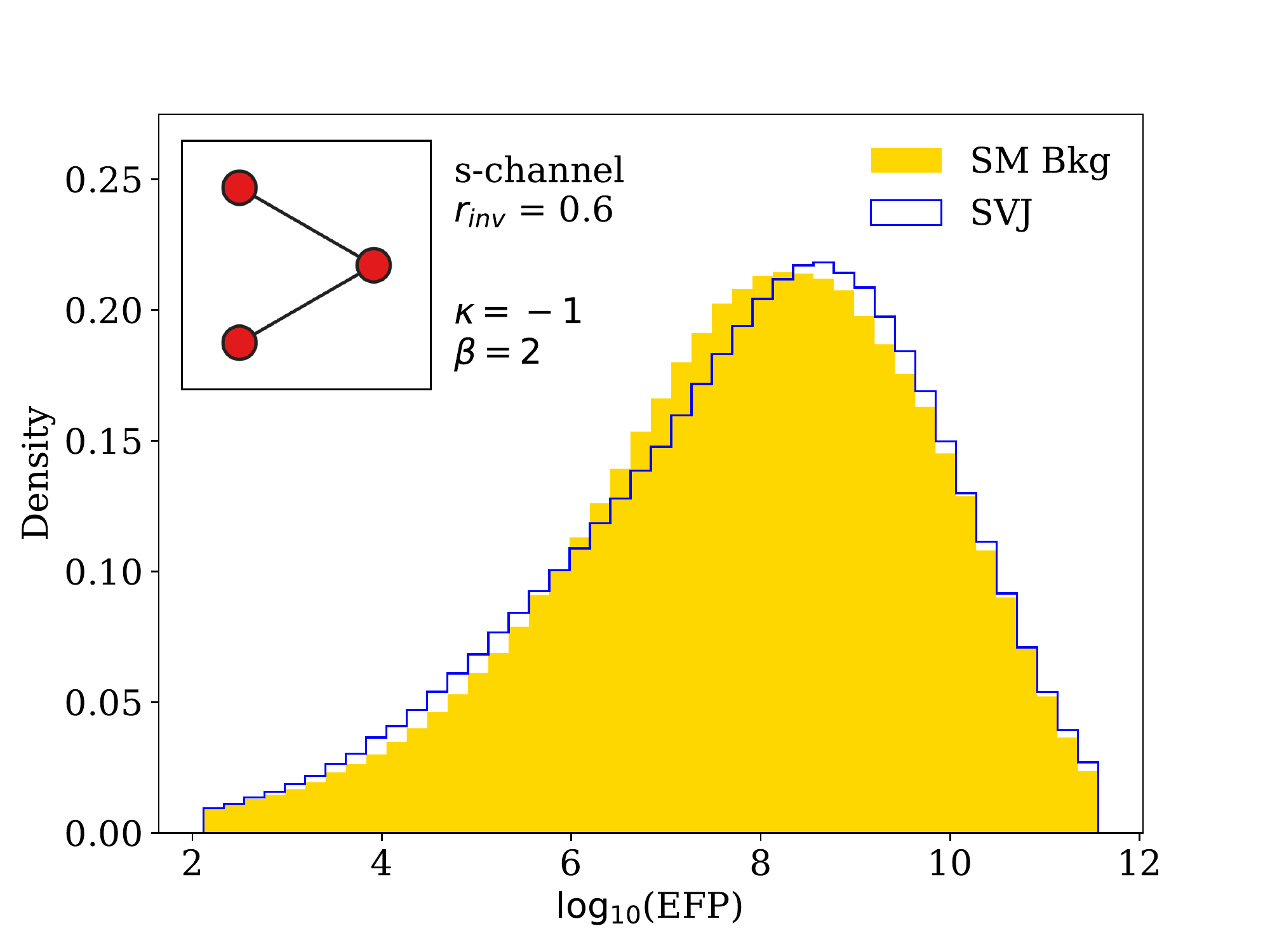}
		\caption{$s$-channel / $\rinv=0.6$}
		\label{fig:graph_s6}
	\end{subfigure}
	\begin{subfigure}[t]{0.32\textwidth}
		\centering
		\includegraphics[width=\textwidth]{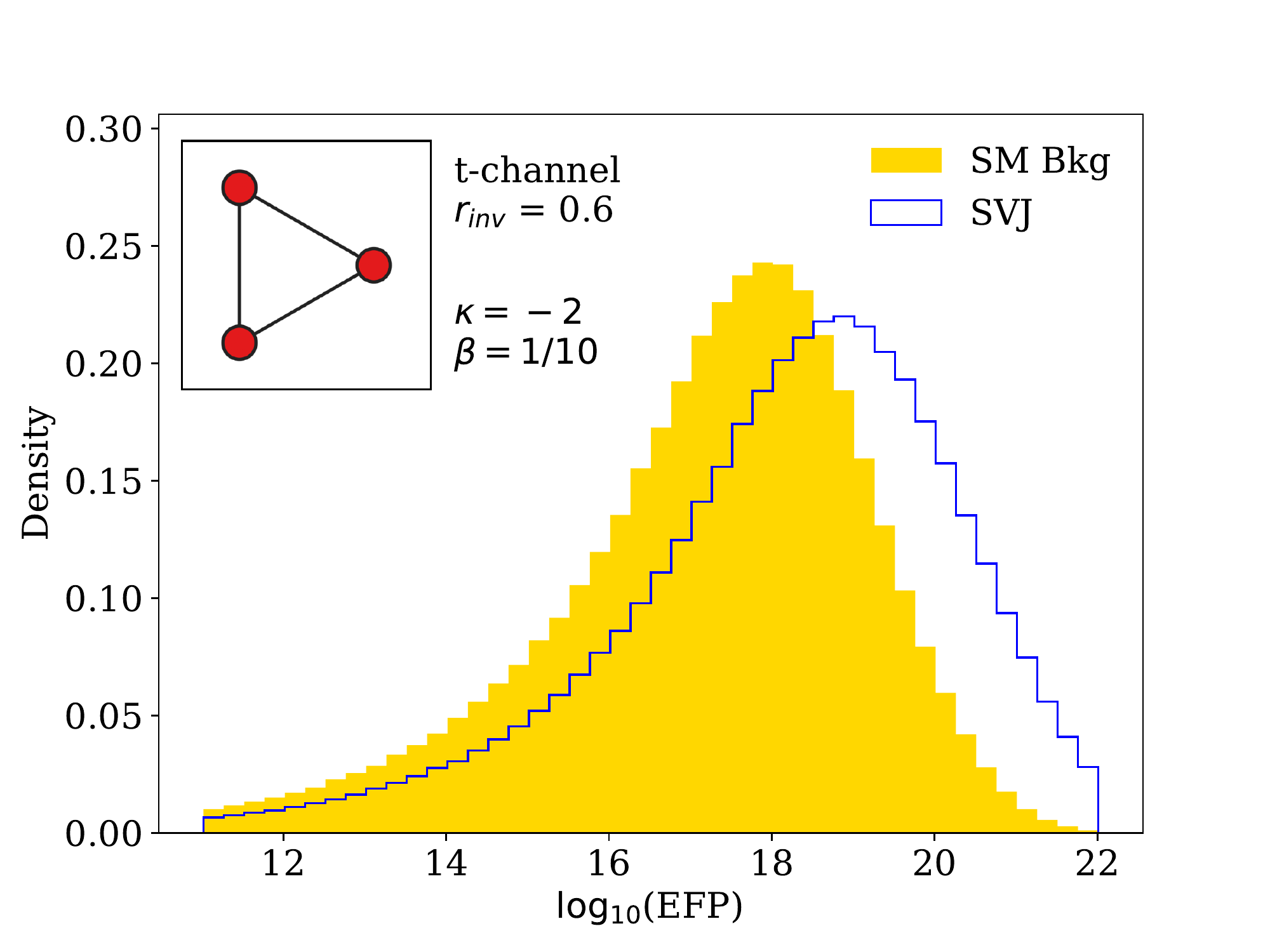}
		\caption{$t$-channel / $\rinv=0.6$}
		\label{fig:graph_t6}
	\end{subfigure}
	\caption{Distribution of  EFP observables selected by the guided search for the $s$-channel $\rinv=0.3$, $\rinv=0.6$ and $t$-channel $\rinv=0.6$ scenarios, shown for semi-visible jets (SVJ) as well as QCD jets from the Standard Model background (SM Bkg). The inset panes show the graph corresponding to the selected EFP. See text for additional details.}
	\label{fig:guided_graphs}
\end{figure*}

In the remaining $s$-channel and $t$-channel examples, addition of the selected EFP improves performance  but fails to match the PFN. Of the three existing gaps, the $s$-channel process with $\rinv=0.3$ is the only result in which we see an IRC-safe EFP observable, where $\kappa=1$. In the other cases, the EFP graphs again have $\kappa<0$, making them sensitive to low-$p_{\textrm{T}}$ information. The complete expression for each selected graph is given in \Eq{eq:guided} and the distributions of each observable for signal and background are shown in \Fig{fig:guided_graphs}.  The triangular graph selected in the case of the $t$-channel process with invisible fraction $\rinv=0.6$ has the same structure as the energy correlation ratio $(e_3)$, though with distinct $\kappa$ and $\beta$ values.

\begin{subequations}
	\label{eq:guided}
	\begin{align}
		\gEq{0.05}{2}{3}{0}{1}{\half}   & = \sum\limits_{a,b=1}^N z_a z_b \left(\theta_{ab}\right)^{3/2} \label{eq:bar3}                                               \\
		\gEq{0.05}{3}{2}{0}{-1}{2}      & = \sum\limits_{a,b,c=1}^N \frac{\left(\theta_{ab} \theta_{ac}\right)^2}{z_a z_b z_c} \label{eq:triangle1}                    \\
		\gEq{0.05}{3}{3}{1}{-2}{\tenth} & = \sum\limits_{a,b,c=1}^N \frac{\left(\theta_{ab} \theta_{bc} \theta_{ac}\right)^{\tenth}}{z_a z_b z_c} \label{eq:triangle2}
	\end{align}
\end{subequations}

\section{Greedy Search}
\label{sec:greedy}
Given the persistent gap between the performance of the LL networks and the HL models augmented by EFP observables, we consider whether the EFP space lacks the needed observables, or whether the guided search is failing to identify it. We examine this question by taking a more comprehensive look at the  space of EFPs we consider. Similar to the technique described in \rref{Faucett2020}, we perform a greedy search in the same EFP space studied in the guided iteration approach explored above. In a greedy search, we explicitly train a new model for each candidate EFP, combining the EFP with the existing 15 HL features. Note that this is significantly more computationally intensive than evaluation of the ADO, as done in the guided search, and seeks to maximize AUC rather than to align decision ordering with the PFN.   The candidate EFP which produces the best-performing model is kept as the 16th HL observable \rev{(Pass 1)}, and the process is repeated in search of a 17th \rev{(Pass 2)}, until a plateau in performance is observed. 

The results of the greedy search across all choices of $\rinv$ in $s$-channel and $t$-channel scenarios where a gap between HL and LL exists are given in \Tab{tab:greedy_results}. Similar levels of performance are achieved as in the guided search, to within statistical uncertainties. In all cases, the HL and LL gap persists. Results are given for the IRC-unsafe selections with dimension $d\leq 3$. A similar greedy search was also performed on the IRC-unsafe $d\leq 5$ EFP set with no performance differences observed.

\begin{table*}
    \caption{ Summary \rev{of two passes} of a greedy search through EFP space for additional observables which might capture the information used by the low-level network (PFN) and match its performance, as measured by AUC.  For each of the four processes and $\rinv$ scenarios in which we have identified a gap between performance of the PFN and the HL model, two passes are made to identify the EFP which most improves the AUC of a new HL model which incorporates the candidate EFP.  Performance of the HL model deduced by the guided search (~\Tab{tab:guided_search_results}) and the PFN are also given. Statistical uncertainty in each case is $\pm 0.002$ with 95\% confidence, measured using bootstrapping over 100 models.
    }
	\label{tab:greedy_results}
	\begin{tabular}{cc|c|cccc|cccc|c|c}
		\toprule
		
		 &  & HL & \multicolumn{4}{c|}{Pass 1} & \multicolumn{4}{c|}{Pass 2} & Guided & PFN  \\
		   Process     & $\rinv$      &  AUC & Graph &  $\kappa$ & $\beta$ & AUC &  Graph &  $\kappa$ & $\beta$ & AUC & HL AUC & AUC\\ \hline
		   $s$-channel & 0.0 & 0.861 & $\gPlot{0.065}{2}{1}{0}$         & $\half$                         & 2                               & 0.864  & $\gPlot{0.065}{2}{2}{0}$         & 2                               & $\tenth$                        & 0.866 & 0.864 & 0.866\\
		  $s$-channel & 0.3  & 0.803 & $\gPlot{0.065}{2}{3}{0}$ & 4        & 2       & 0.807 & $\gPlot{0.065}{2}{1}{0}$ & -1       & 1       & 0.809 & 0.807 & 0.822\\
		   $s$-channel & 0.6 & 0.736 & $\gPlot{0.065}{2}{2}{0}$ & 4        & 4        & 0.744 & $\gPlot{0.065}{2}{3}{0}$ & -2       & $\tenth$ & 0.747 & 0.747 & 0.776 \\
		   $t$-channel & 0.6 & 0.683 & $\gPlot{0.065}{3}{2}{0}$ & -1       & $\tenth$ & 0.690 &   $\gPlot{0.065}{4}{3}{1}$ & -2       & 4        & 0.692 & 0.690 & 0.697\\
		   \bottomrule
	\end{tabular}
\end{table*}

The greedy search selects similar EFP graphs as the guided search, with the exception of the 4-node graph selected in pass 2 for the $\rinv=0.6$, $t$-chanel scenario. No IRC-safe graphs ($\kappa=1$) are selected and we again see frequent sensitivity to low $p_{\textrm{T}}$ parameters (i.e. $\kappa=-2, -1, \half$) and a variety of both narrow and wide angle features (i.e. $\beta = \tenth, 2, 4$). A repeat of the greedy search with only IRC-safe observables achieves no performance improvements over the original HL features, suggesting that missing information may be strongly tied to IRC-unsafe representations of the information in the jet constituents.

\section{Exploration of $p_{\textrm{T}}$ Dependence}
\label{sec:low_pt}
\begin{figure}
    \centering
    \includegraphics[width=0.45\textwidth]{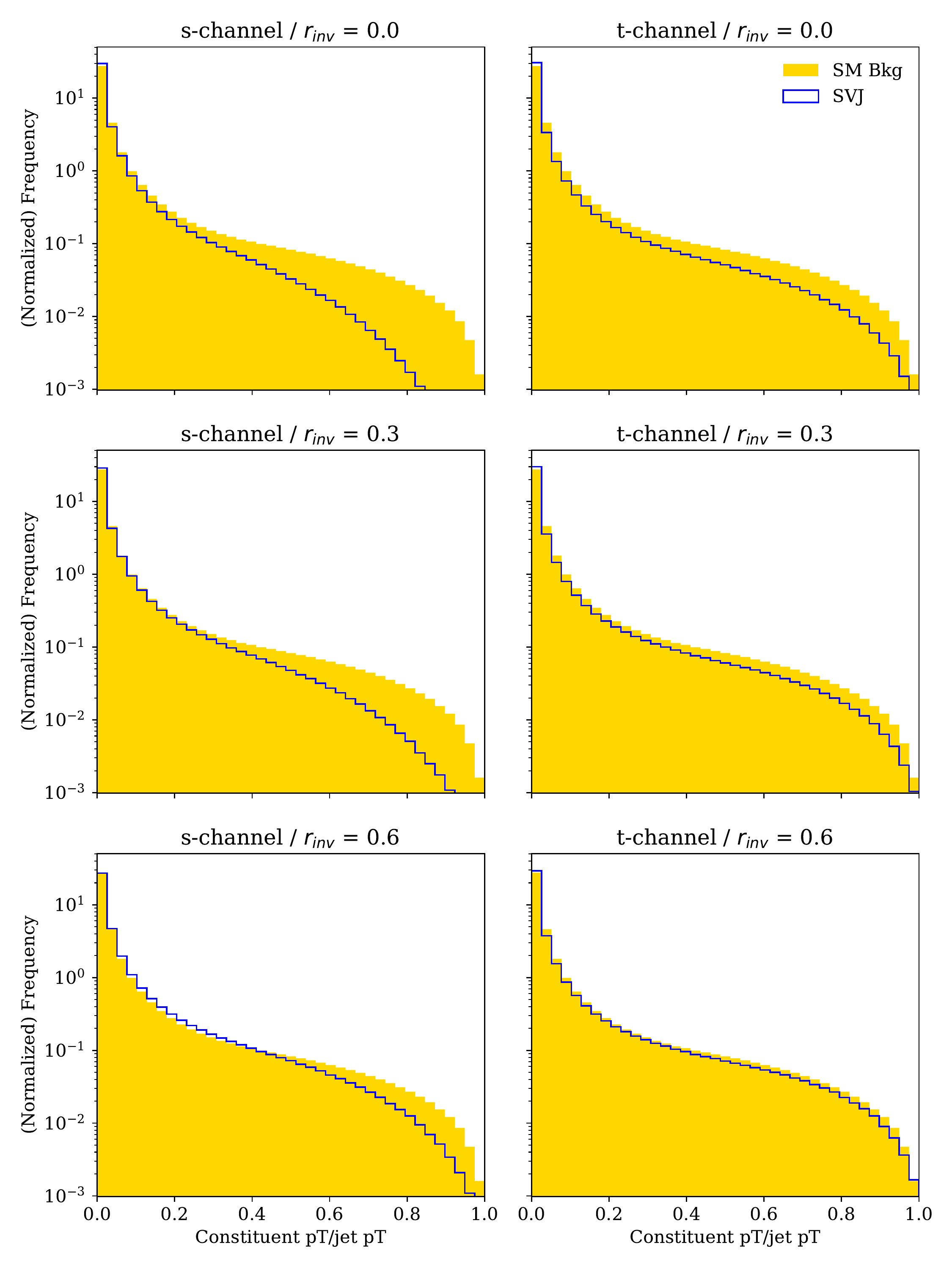}
    \caption{Distribution of constituent $p_{\textrm{T}}$ relative to the jet $p_{\textrm{T}}$ for the six scenarios: $s$-channel and $t$-channel with $\rinv \in [0,0.3,0.6]$.}
    \label{fig:ptfrac}
\end{figure}

The results of both the guided search and the greedy search strongly suggest that the full performance gap which persists between the HL and LL representation of the jet contents cannot be compactly expressed in terms of a small number of EFP observables in the set that have been considered. This raises the question of what feature of the LL constituents can be credited with this performance improvement and why that information does not translate compactly to our EFP observables. The clues from the guided and greedy search point to sensitivity to low-$p_{\textrm{T}}$ constituents, as can be generated from soft radiation.  Figure~\ref{fig:ptfrac} shows the distribution of constituent $p_{\textrm{T}}$ relative to the jet $p_{\textrm{T}}$ for the six scenarios, in which SVJs  appear to have more constituents at low relative $p_{\textrm{T}}$. Recall that jet constituents are trimmed if the subjet they belong to has a $p_{\textrm{T}}$ below a threshold of 5\% of the $p_{\textrm{T}}$ of the jet\rev{, corresponding to $f_\textrm{cut}=0.05$.}

To consider whether the distinguishing information is contained in these low-$p_{\textrm{T}}$ constituents, \rev{we explore a broader range of thresholds, both lowering it to 0\% and raising it  to 10\% and 15\%}. The HL features are re-evaluated on the newly trimmed constituents, and used as inputs to a new LightGBM model, whose performance is compared to a PFN trained on the trimmed constituents. 
Results for networks \rev{with varying $f_\textrm{cut}$ thresholds} are shown in \Tab{tab:low_pt_cut}.

\rev{
\begin{table*}
	\centering
	\caption{ Comparison of the performance difference between a PFN  operating on low-level constituent information and a LightGBM model using high-level summary quantities, for several values of the jet trimming parameter $f_\textrm{cut}$. Jet constituents belonging to a subjet whose fraction of the jet $p_\textrm{T}$ is below $f_\textrm{cut}$ are dropped, which has the effect of removing lower-$p_\textrm{T}$ constituents. Shown is the AUC of each model for $s$- and $t$-channel processes under three $\rinv$ scenarios. Statistical uncertainty in each case is $\pm 0.002$ with 95\% confidence, measured using bootstrapping over 100 models.}
	\label{tab:low_pt_cut}
	\begin{tabular}{c|ccc|ccc|ccc}
		\toprule
		\multicolumn{10}{c}{$s$-channel}\\
		              & \multicolumn{3}{c}{$\rinv= 0.0$} & \multicolumn{3}{c}{$\rinv= 0.3$} & \multicolumn{3}{c}{$\rinv= 0.6$}                                 \\ 
		$f_\textrm{cut}$ & PFN                          & LightGBM     & LL-HL Gap                    & PFN                          & LightGBM & LL-HL Gap & PFN & LightGBM & LL-HL Gap \\ \midrule
        0.00    & 0.908 & 0.895 & 0.013 & 0.853 & 0.829 & 0.024 & 0.788  & 0.739 & 0.049  \\
		0.05    & 0.866                            & 0.861  & 0.005                          & 0.822                            & 0.803 & 0.019   & 0.776   & 0.736   & 0.040  \\
		0.10   & 0.847                            & 0.848  & -0.001                          & 0.790                            & 0.790  & 0.000   & 0.746   & 0.721   & 0.025 \\
		0.15   & 0.838                            & 0.843 & -0.005                           & 0.784                            & 0.785   & -0.001  & 0.738   & 0.717  & 0.021   \\ \bottomrule
		\multicolumn{7}{c}{\vspace{0.5em}}\\ \toprule
        \multicolumn{10}{c}{$t$-channel}\\
		              & \multicolumn{3}{c}{$\rinv= 0.0$} & \multicolumn{3}{c}{$\rinv= 0.3$} & \multicolumn{3}{c}{$\rinv= 0.6$}                                 \\ 
		$f_\textrm{cut}$ & PFN                          & LightGBM   & LL-HL Gap                       & PFN                          & LightGBM & LL-HL Gap & PFN & LightGBM & LL-HL Gap \\ \midrule
		0.00    & 0.825 & 0.817 & 0.008 & 0.748 & 0.737 & 0.011 & 0.662  & 0.647 & 0.0015  \\
		0.05   & 0.806                            & 0.808     & -0.002                       & 0.754                            & 0.755  & -0.001  & 0.697   & 0.683  & 0.014  \\
		0.10   & 0.741                            & 0.742     & -0.001                       & 0.662                            & 0.663    & -0.001& 0.595   & 0.597  & -0.002  \\
		0.15  & 0.731                            & 0.740     & -0.009                      & 0.655                            & 0.661    & -0.006 & 0.593   & 0.596 & -0.003    \\ \bottomrule
	\end{tabular}
\end{table*}
}
\rev{
In each case, raising the  $f_\textrm{cut}$ threshold decreases the classification performance, as might be expected due to the removal of low-$p_\textrm{T}$ information. Perhaps more interesting is the variation in the gap between the performance of the HL LightGBM model and the PFN  operating on low-level constituents. In nearly every case, the gap grows as more low-$p_\textrm{T}$ information is included, supporting the hypothesis that this is the origin of most of the information missing from the HL models.  The details of the low-$p_\textrm{T}$ constituents are likely to be very sensitive to modeling uncertainties and subject to concerns about infrared and collinear safety. 

However, even in for the most aggressive value of $f_\textrm{cut}=0.15$, a persistent gap of $\Delta$AUC=0.021 remains in the $s$-channel, $r_\textrm{inv}=0.6$ scenario, which cannot be explained by low-$p_\textrm{T}$ constituents. We therefore examine this in more detail.

  First, we consider the EFPs selected in the study above where $f_\textrm{cut}=0.05$, including EFPs selected by the guided search (\Tab{tab:guided_search_results}) and the greedy search (\Tab{tab:greedy_results}).  For all combinations of HL and identified EFPs, no performance gain is seen.  Next, we perform a fresh guided search on models trained from the $f_\textrm{cut}=0.15$ constituents.  In contrast to previous cases, no large improvement in training performance is obtained on the first selected EFP. After 200 iterations, the gap is reduced to $\Delta$AUC=0.010 with the addition of 200 EFPs. We note that this far exceeds the mean number of constituents of these trimmed jets, 60.  We conclude that there there does not appear to be a compact representation of the remaining information in the EFP space we have explored.
}

\section{Conclusions}
\label{sec:discussion}

We have analyzed the classification performance of models trained to distinguish background jets from semi-visible jets using the low-level jet constituents, and found them to offer stronger performance than models which rely on high-level quantities motivated by other processes, mostly those involving collimated hadronic decays of massive objects.  

While models operating on the existing suite of HL quantities nearly match the performance of those using LL information, a significant gap exists which suggests that relevant information remains to be captured, perhaps in new high-level features.   To our knowledge, this is the first study to compare the performance of constituent-based and high-level semi-visible-jet taggers, and to identify the existence of relevant information uncaptured by existing high-level features. Jets due to semi-visible decays are markedly different in energy distribution than those from massive objects, so it is not unexpected that existing features may not completely summarize the useful information.

Using a guided strategy, we identify a small set of new useful features from the space of energy-flow polynomials, but these do not succeed in completely closing the performance gap.  In most cases, the remaining gap seems to be due to information contained in very low-$p_\textrm{T}$ constituents, which is likely to be sensitive to modeling of showering and hadronization and may not be infrared and collinear safe. This highlights the importance of interpretation and validation of information used by constituent-based taggers. \rvv{ As demonstrated by Ref.~\cite{Cohen:2020afv}, the specific pattern of energy deposition may depend sensitively on both the parameters of the theoretical model as well as the settings chosen for the hadronization model. It is therefore vital that the information being analyzed be interpreted before being applied to analysis of collider data.}

In one case studied here, a gap persist between low- and high-level-based models even when low-$p_\textrm{T}$ constituents are aggressively trimmed, suggesting the possibility that a new high-level feature could be crafted to capture this useful high-$p_\textrm{T}$ information.  Our efforts to capture this information with the simpler energy-flow polynomials was not successful, suggesting that more complex \rvv{high-$p_\textrm{T}$} observables may exist which provide useful discrimination between QCD and semi-visible jets. The studies presented here can inform and guide theoretical work to construct such observables specifically tailored to this category of jets.  \rvv{Whether such observables can be efficiently represented using alternative basis sets of observables, and whether they are robust to \cite{Shimmin:2017mfk} or explicitly dependent on~\cite{Ghosh:2021roe} uncertainties while providing power over large regions of theoretical parameter space is an important avenue for future work.}




\section{Acknowledgements}
\label{sec:acknowledgements}
The authors thank Po-Jen Cheng for his assistance in sample validation. This material is based upon research supported by the Chateaubriand Fellowship of the Office for Science \& Technology of the Embassy of France in the United States.  T. Faucett and D. Whiteson are supported by the U.S.\ Department of Energy (DOE), Office of Science under Grant No. DE-SC0009920. S.-C. Hsu is supported by the National Science Foundation under Grant No. 2110963.

~\newpage
\appendix

\section{Signal reweighting}
\label{app:reweight}

\begin{figure*}
	\includegraphics[width=0.49\textwidth]{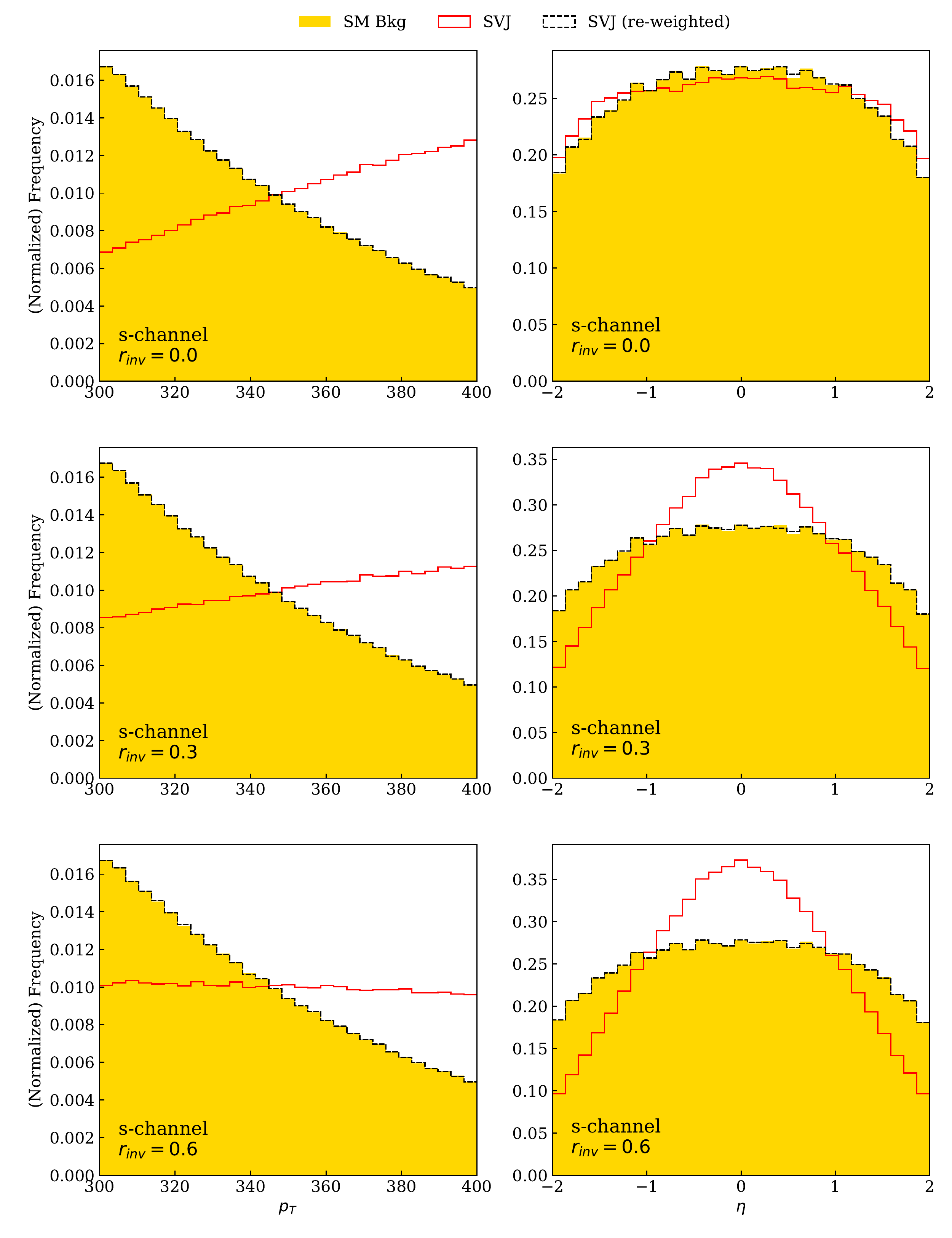}
	\includegraphics[width=0.49\textwidth]{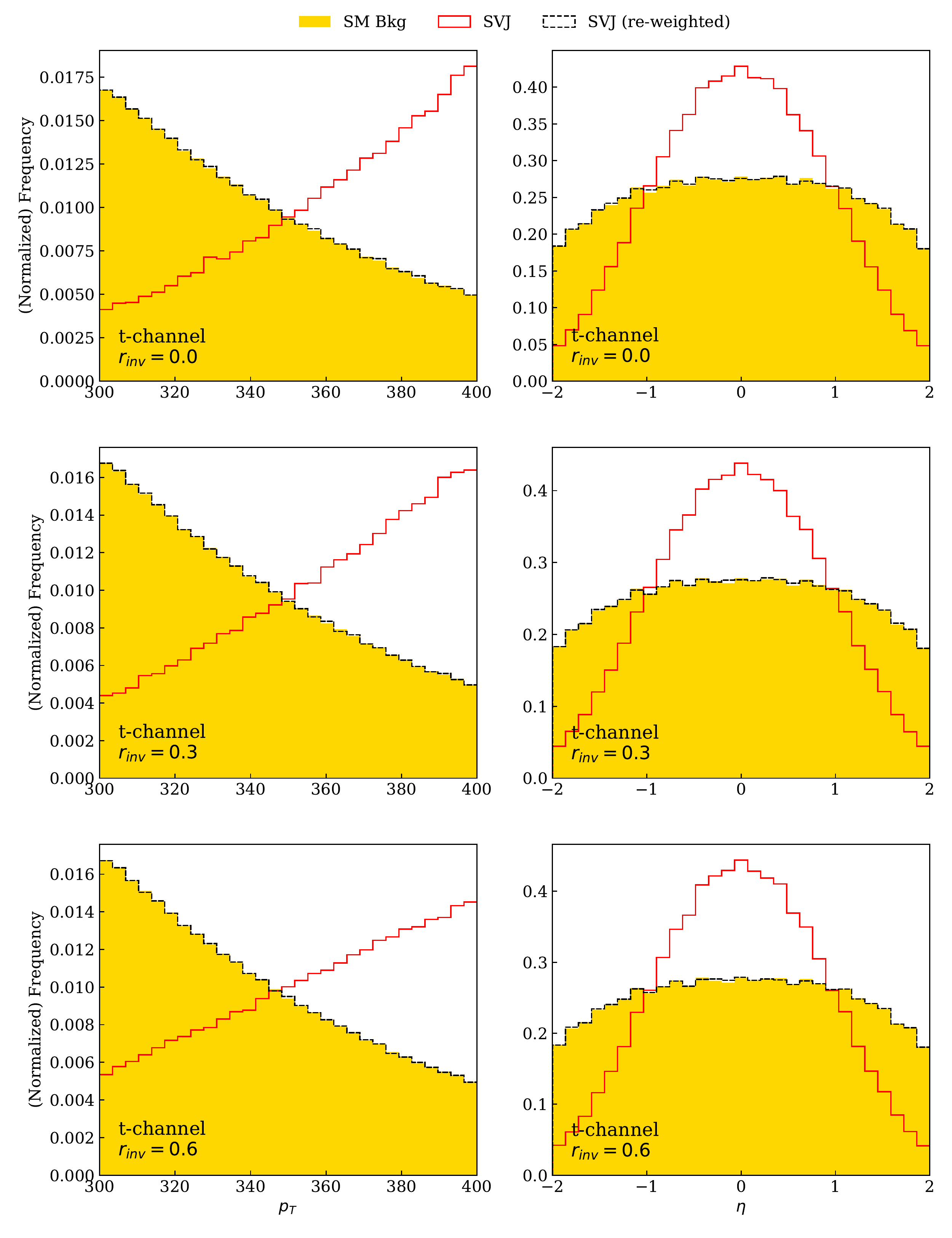}
	\caption{ Distributions of jet transverse momentum ($p_{\textrm{T}}$) and pseudo-rapidity ($\eta$) shown for semi-visible jet (SVJ) and the standard model background (SM Bkg) for the six simulated scenarios, three choices of invisible fraction $\rinv$ for both the $s$-channel and $t$-channel processes. The SVJ samples are reweighted to match the background distributions.}
	\label{fig:reweight}
\end{figure*}

The signal and background distributions have distinct transverse momentum and pseudo-rapidity distributions due to the processes used to generate them. We wish to learn to classify the signal and background independent of these quantities, and so reweight the signal events to match the background distribution, see Fig~\ref{fig:reweight}.

\section{Jet Substructure Observables}
\label{app:jssdescriptions}

High-level features used to discriminate between semi-visible and background jets are defined below.

\subsection{Jet transverse momentum and mass}
\label{app:jss_hist}
The sum of jet $p_{\textrm{T}}$ constituents is included as a HL observable both in the initial HL inputs and along with EFPs to give ML algorithms a relative scale for dimensionless EFP features to train with. The jet $p_{\textrm{T}}$ sum is calculated by
\begin{equation}
	p_{\textrm{T}} = \sum\limits_{i \in \text{jet}} \pti{i}
	\label{eq:pt}
\end{equation}

\begin{figure*}[ht]
	\includegraphics[width=0.48\textwidth]{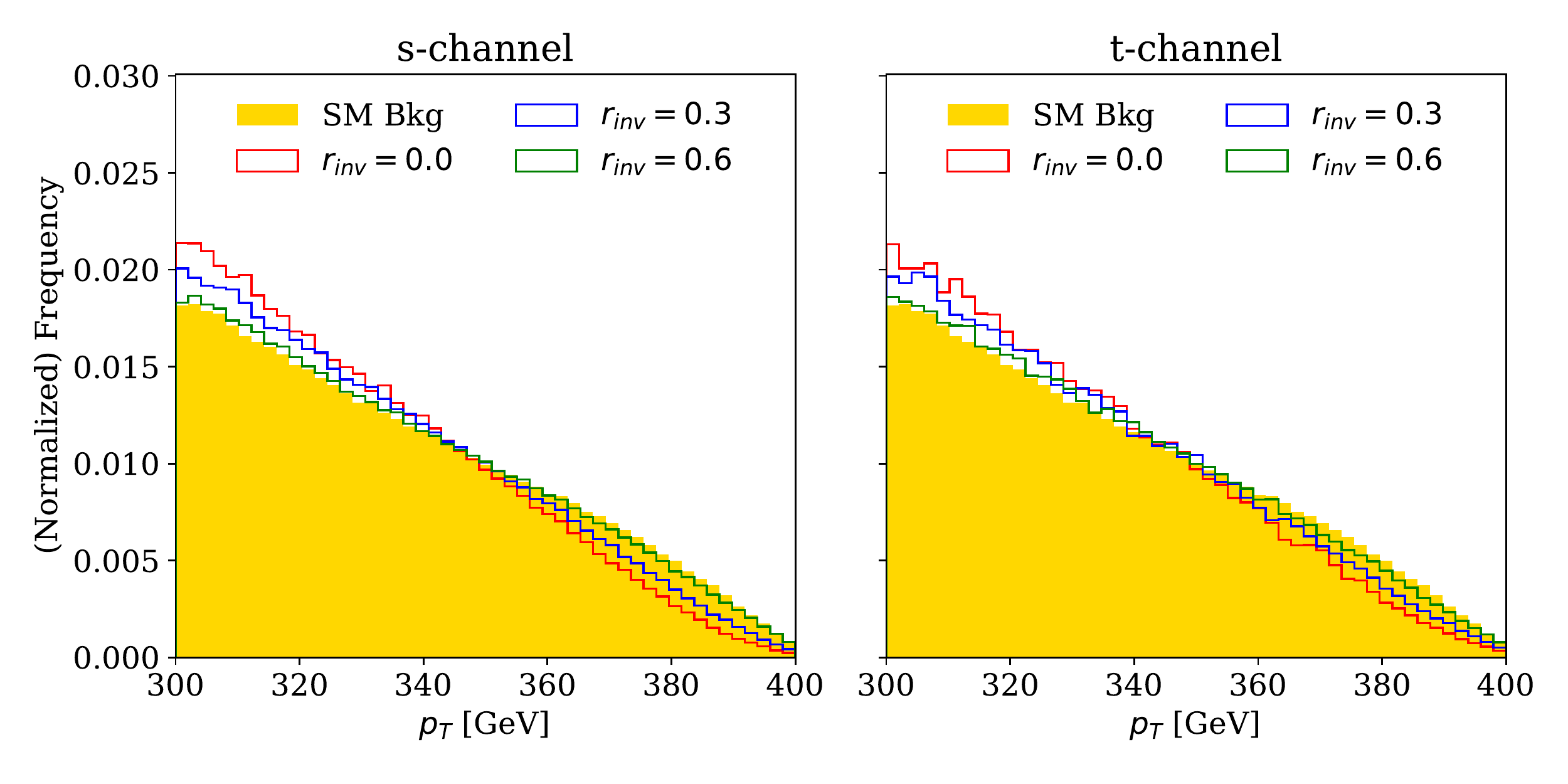}
		\includegraphics[width=0.48\textwidth]{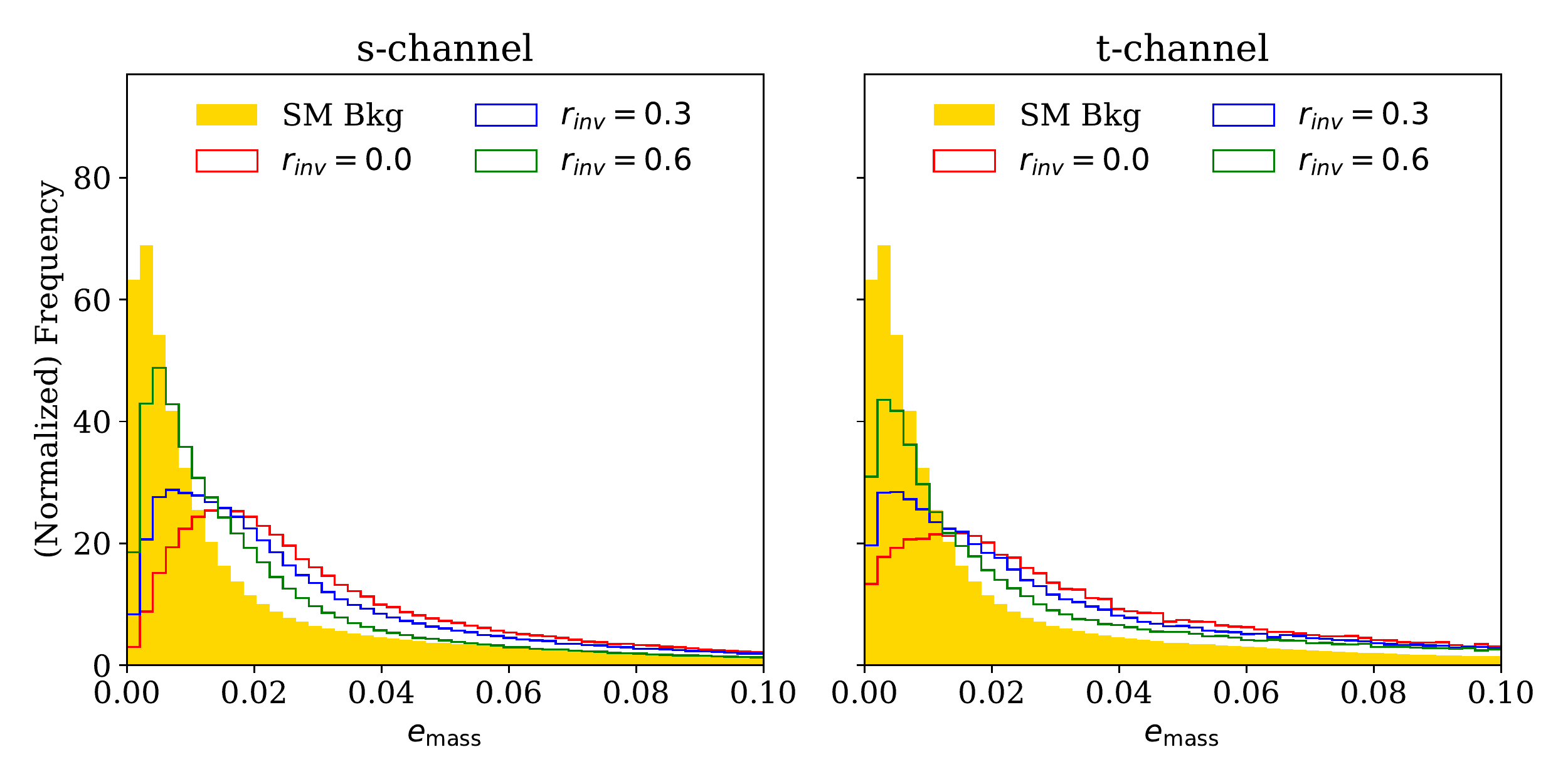}
	\caption{Distributions of jet transverse momentum($p_{\textrm{T}}$) and $e_{\textrm{mass}}$ shown for semi-visible jet and the standard model background (SM Bkg) for the six simulated scenarios, three choices of invisible fraction $\rinv$ for both the $s$-channel and $t$-channel processes.}
	\label{fig:mass_pt}
\end{figure*}

Distributions for jet $p_{\textrm{T}}$ and $e_\textrm{mass}$, defined below, are shown in \Fig{fig:mass_pt}.

\subsection{Generalized Angularities}
\begin{figure*}[ht]
	\centering
	\includegraphics[width=0.48\textwidth]{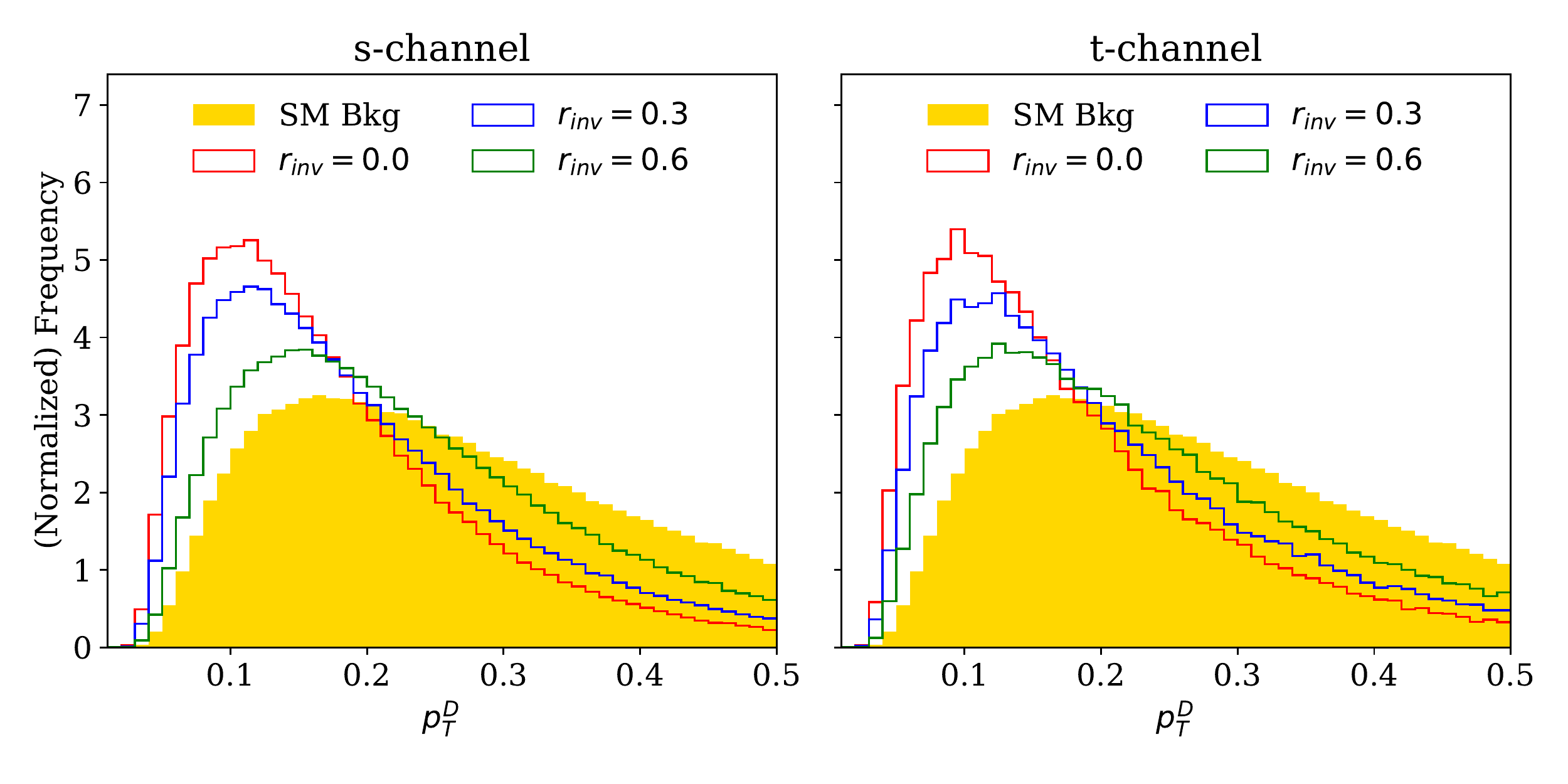}
	\includegraphics[width=0.48\textwidth]{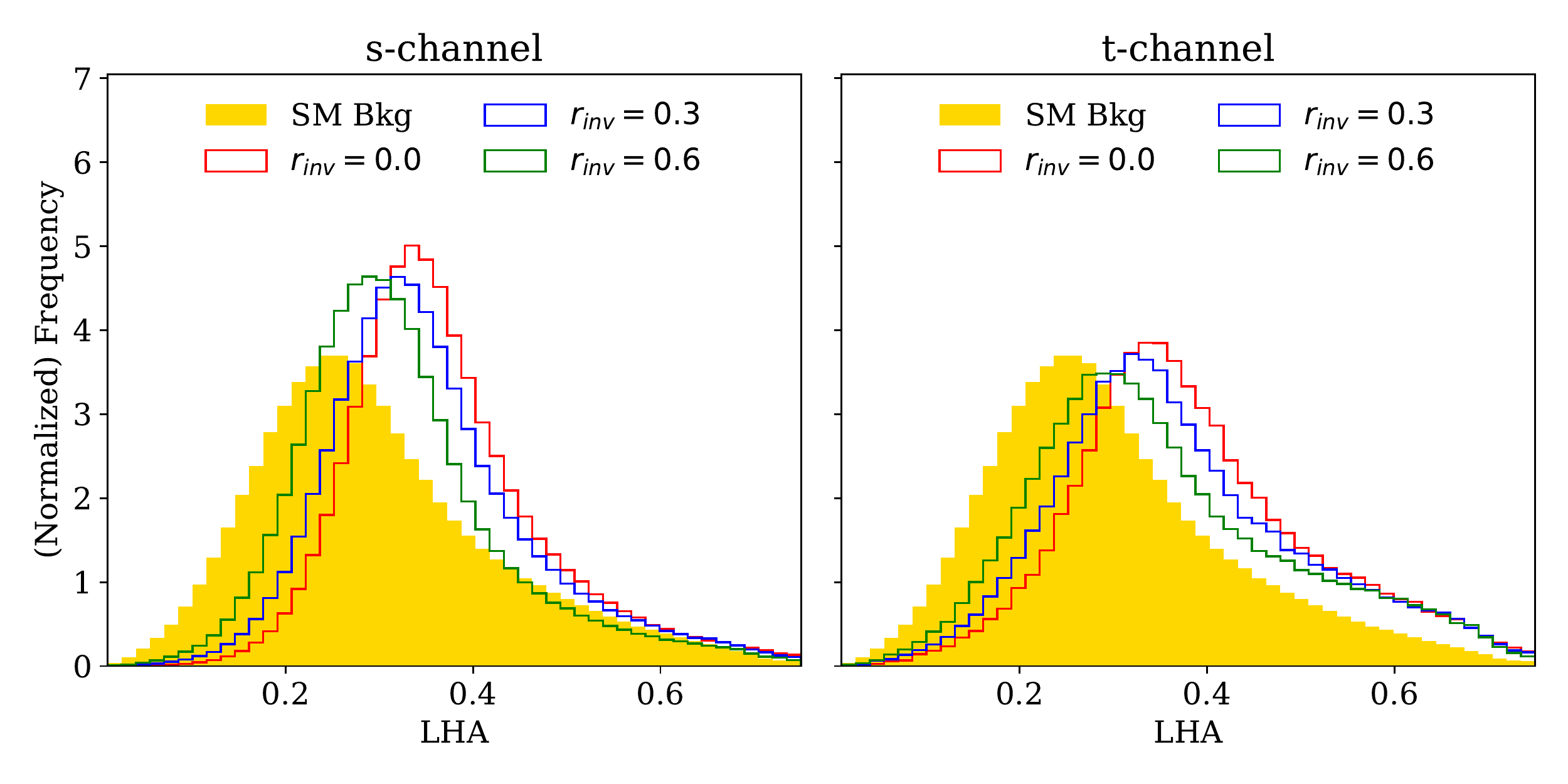}
	\includegraphics[width=0.48\textwidth]{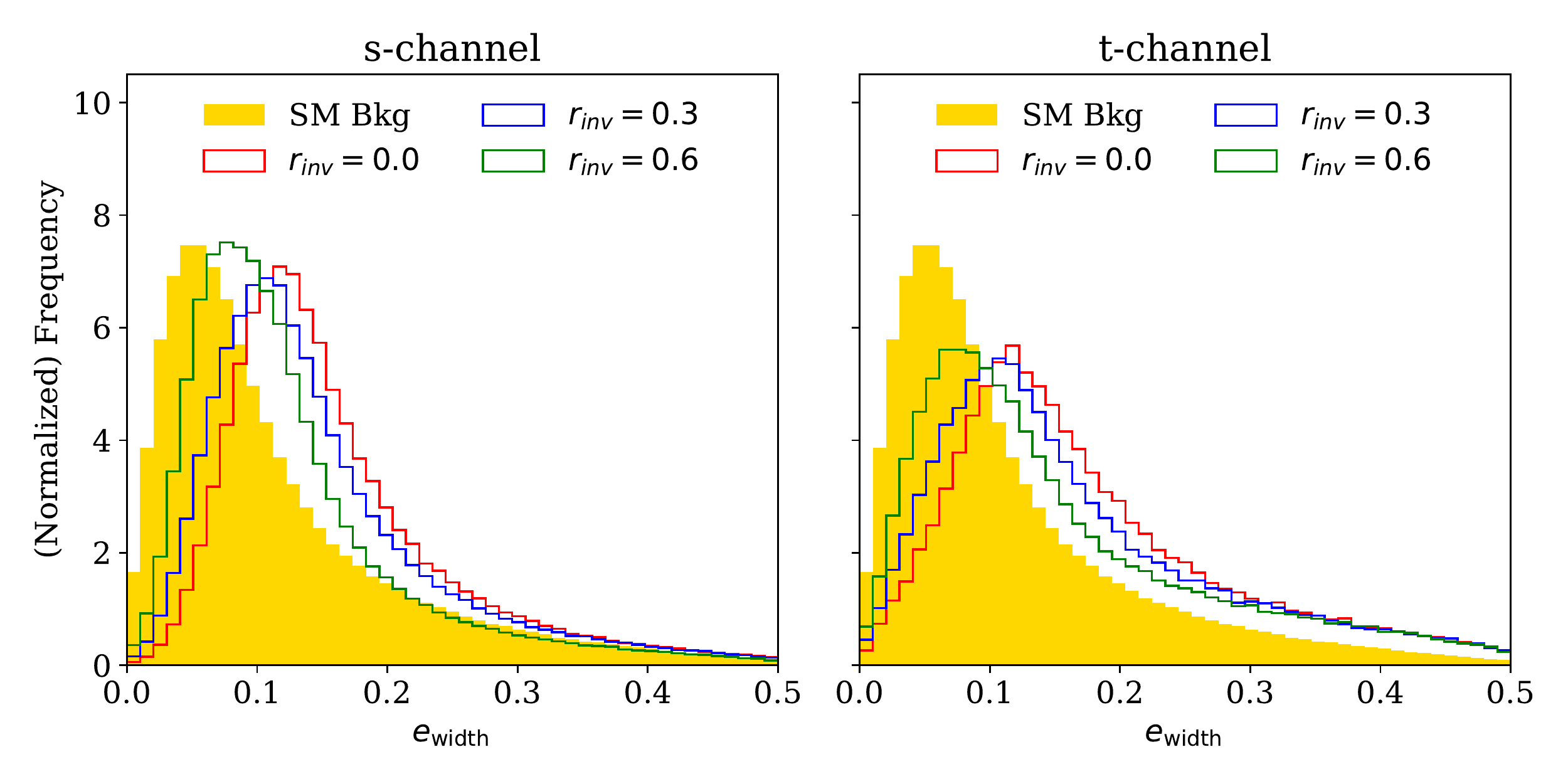}
	\includegraphics[width=0.48\textwidth]{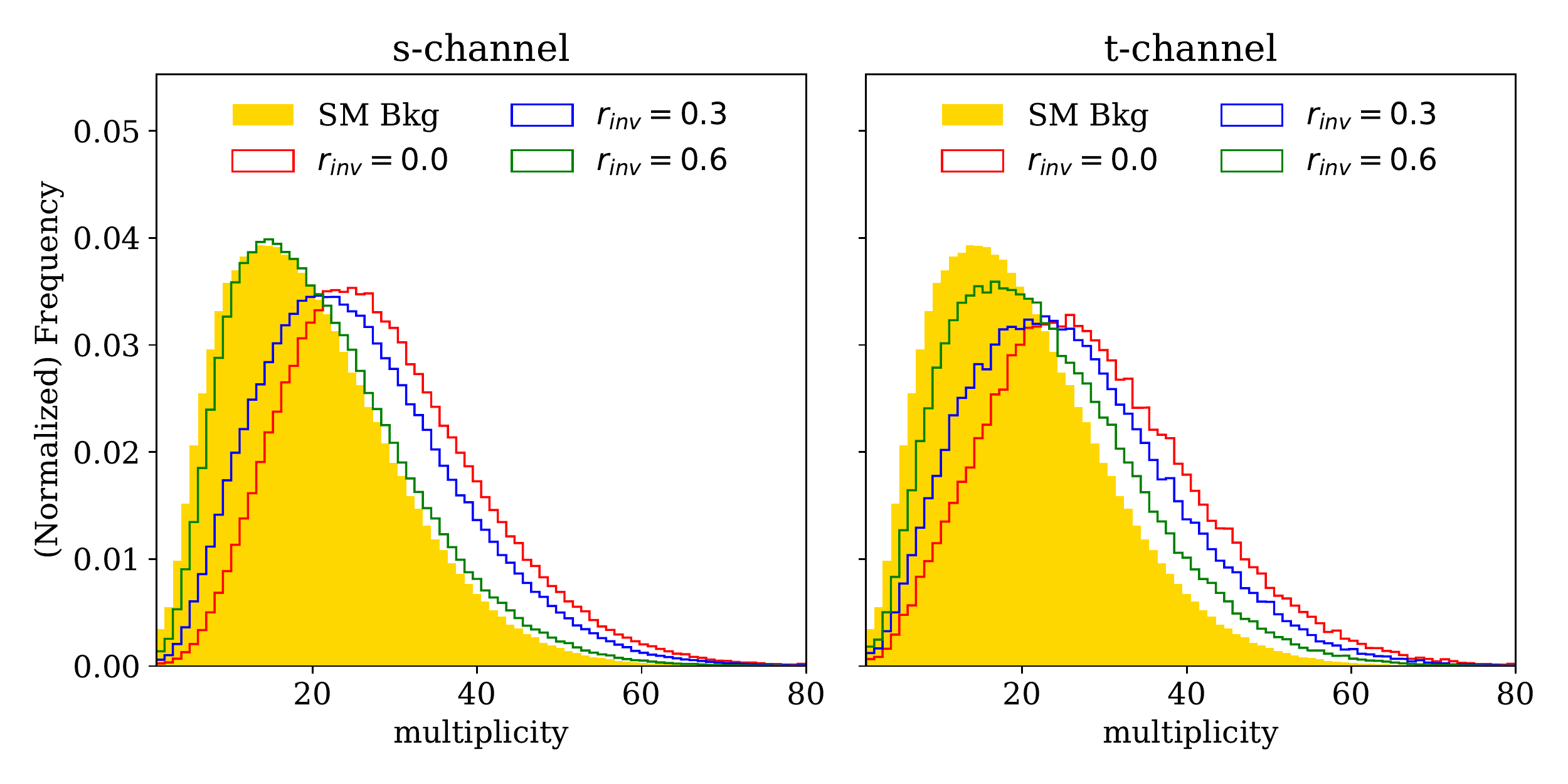}
	\caption{ Distributions of jet $p_{\textrm{T}}^D$, $\textrm{LHA}$, $\textrm{width}$ and multiplicity shown for semi-visible jet (red, green, blue) and the standard model background (SM Bkg, yellow) for the six simulated scenarios, three choices of invisible fraction $\rinv$ for both the $s$-channel and $t$-channel processes.}
	\label{fig:ga}
\end{figure*}

Multiple standard HL observables are defined by choices of $\kappa$ and $\beta$ parameters from the momentum fraction $\left(z_i\right)$ and angular separation $\left(\theta_{i}\right)$ of a Generalized Angularity (GA) expression~\cite{Larkoski:2014vc},
\begin{equation}
	\lambda_{\beta}^{\kappa} = \sum\limits_{i \in \text{jet}} z_i^{\kappa} \theta_i^{\beta}
\end{equation}
The Les Houches Angularity ($\textrm{LHA}$) is defined from the GA expression with parameters $\kb{1}{1/2}$ and $p_{\textrm{T}}^D$ with $\kb{2}{0}$. Written explicitly, these become
\begin{align}
	\textrm{LHA}     & = \sum\limits_{i \in \text{jet}} z_i \theta_i^{1/2} \\
	p_{\textrm{T}}^D & = \sum\limits_{i \in \text{jet}} z_i^2
\end{align}
Two additional values, $e_\textrm{width}$ and $e_\textrm{mass}$, are produced by choices of $\kb{1}{1}$ and $\kb{1}{2}$, respectively
\begin{align}
	e_\textrm{width} & = \sum\limits_{i \in \text{jet}} z_i \theta_i   \\
	e_\textrm{mass}  & = \sum\limits_{i \in \text{jet}} z_i \theta_i^2
\end{align}
Lastly, the multiplicity (although technically defined as simply the total number of constituents in the jet) can be expressed in this same generalized form for $\kb{0}{0}$
\begin{equation}
	\text{multiplicity} = \sum\limits_{i \in \text{jet}}\, 1
\end{equation}
Distributions for all GA observables are shown in \Fig{fig:ga}

\subsection{Energy Correlation}
\begin{figure*}[ht]
	\centering
	\includegraphics[width=0.48\textwidth]{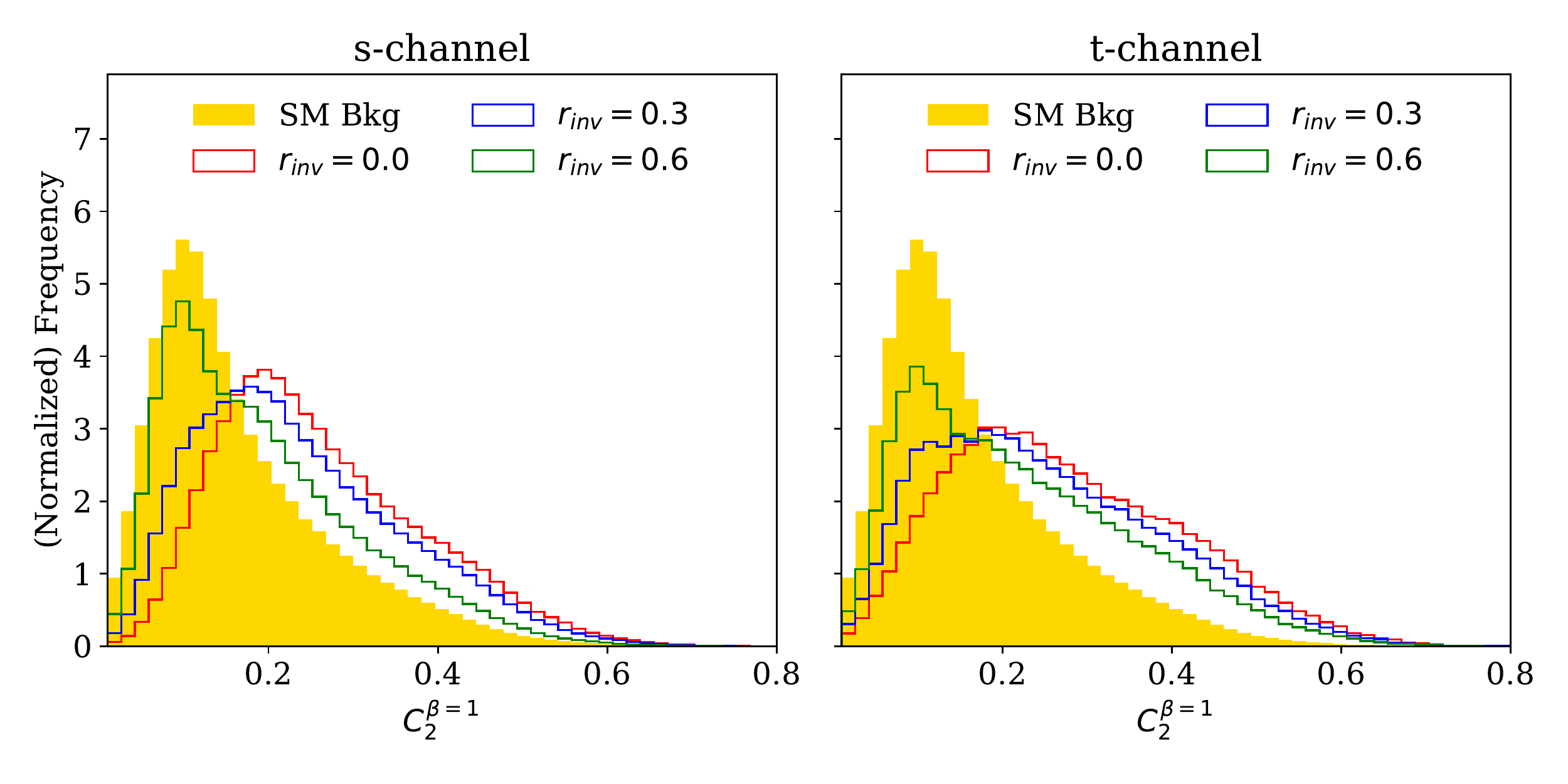}
	\includegraphics[width=0.48\textwidth]{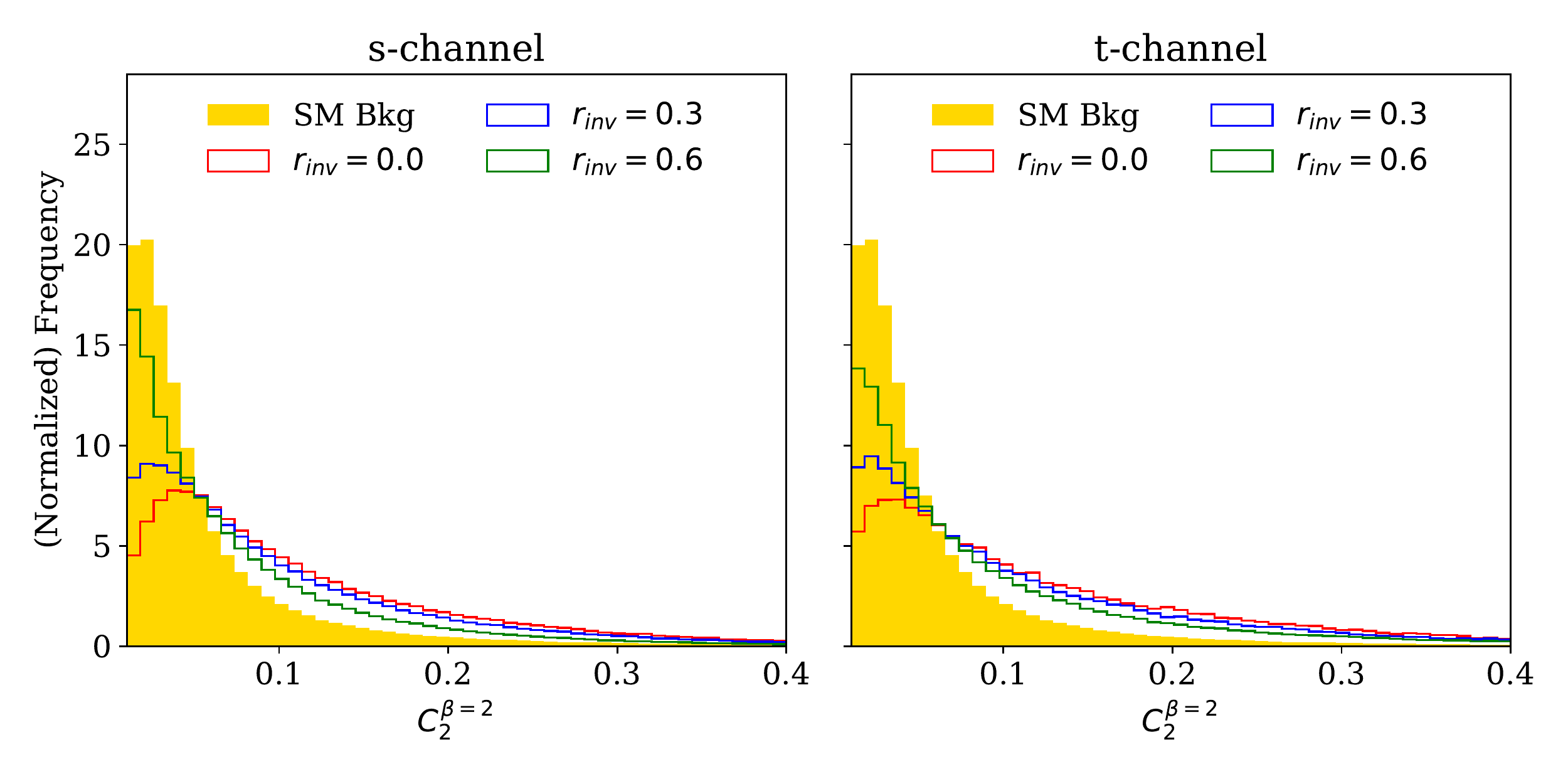}
	\includegraphics[width=0.48\textwidth]{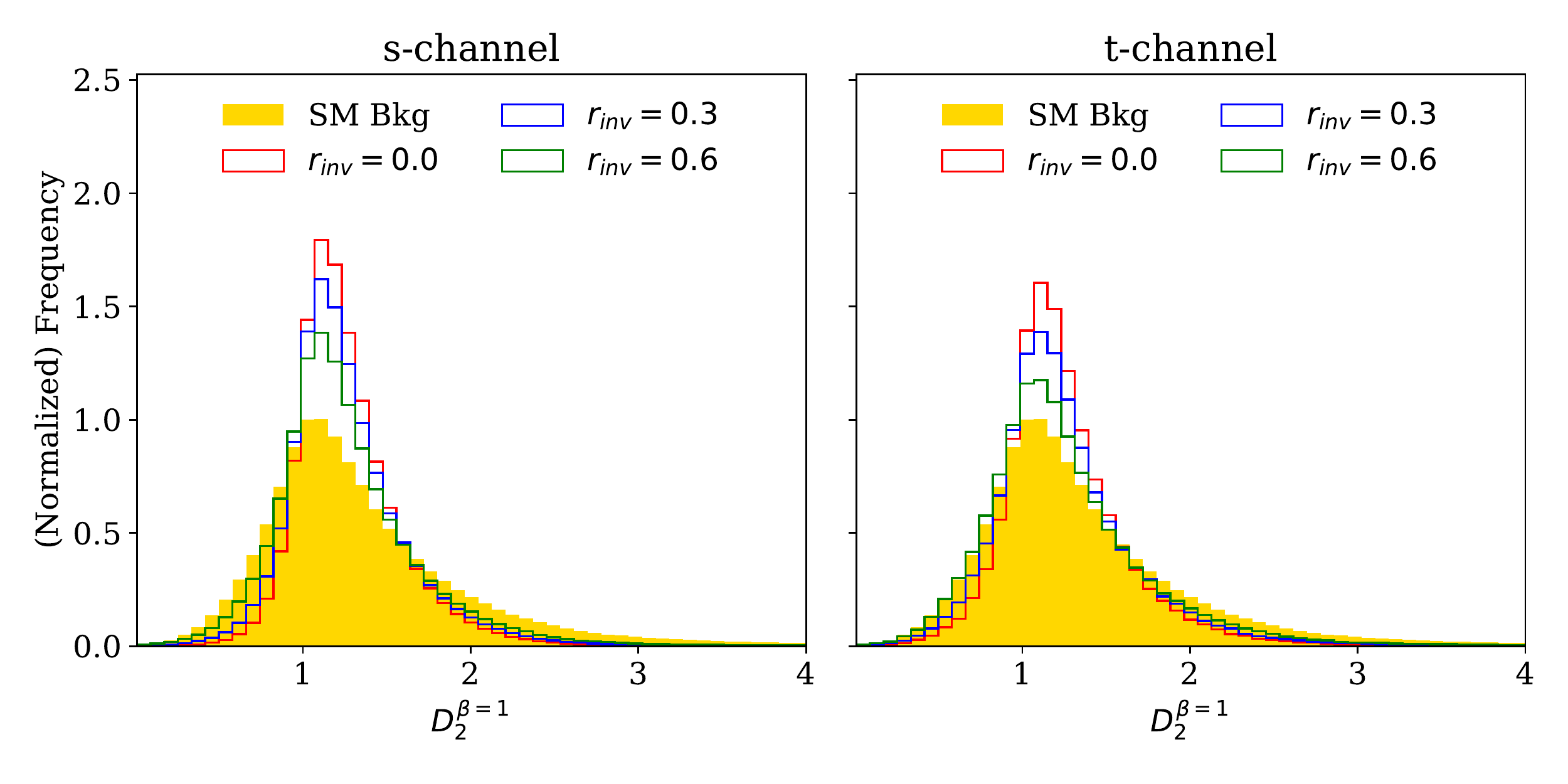}
	\includegraphics[width=0.48\textwidth]{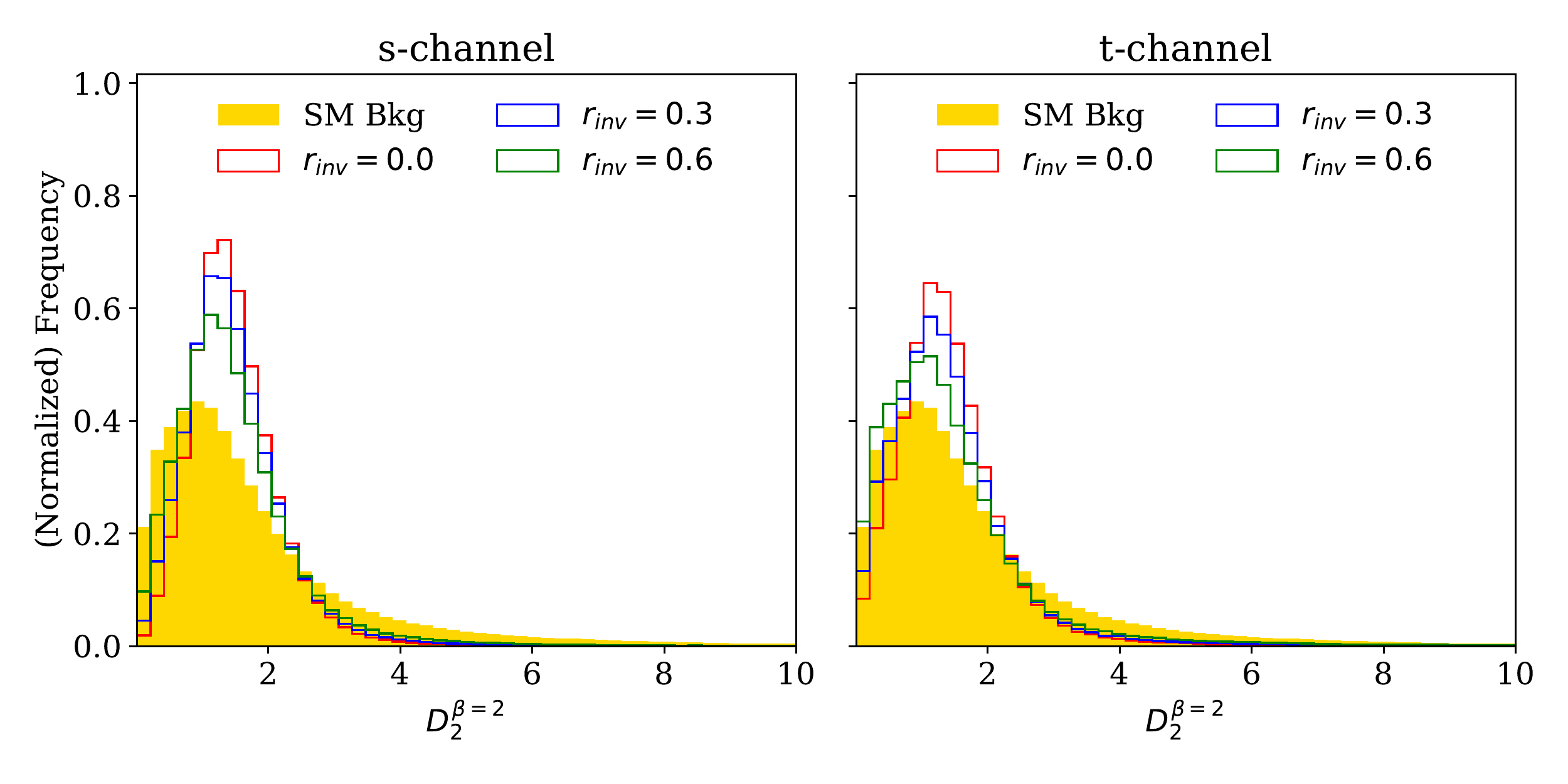}
	\includegraphics[width=0.48\textwidth]{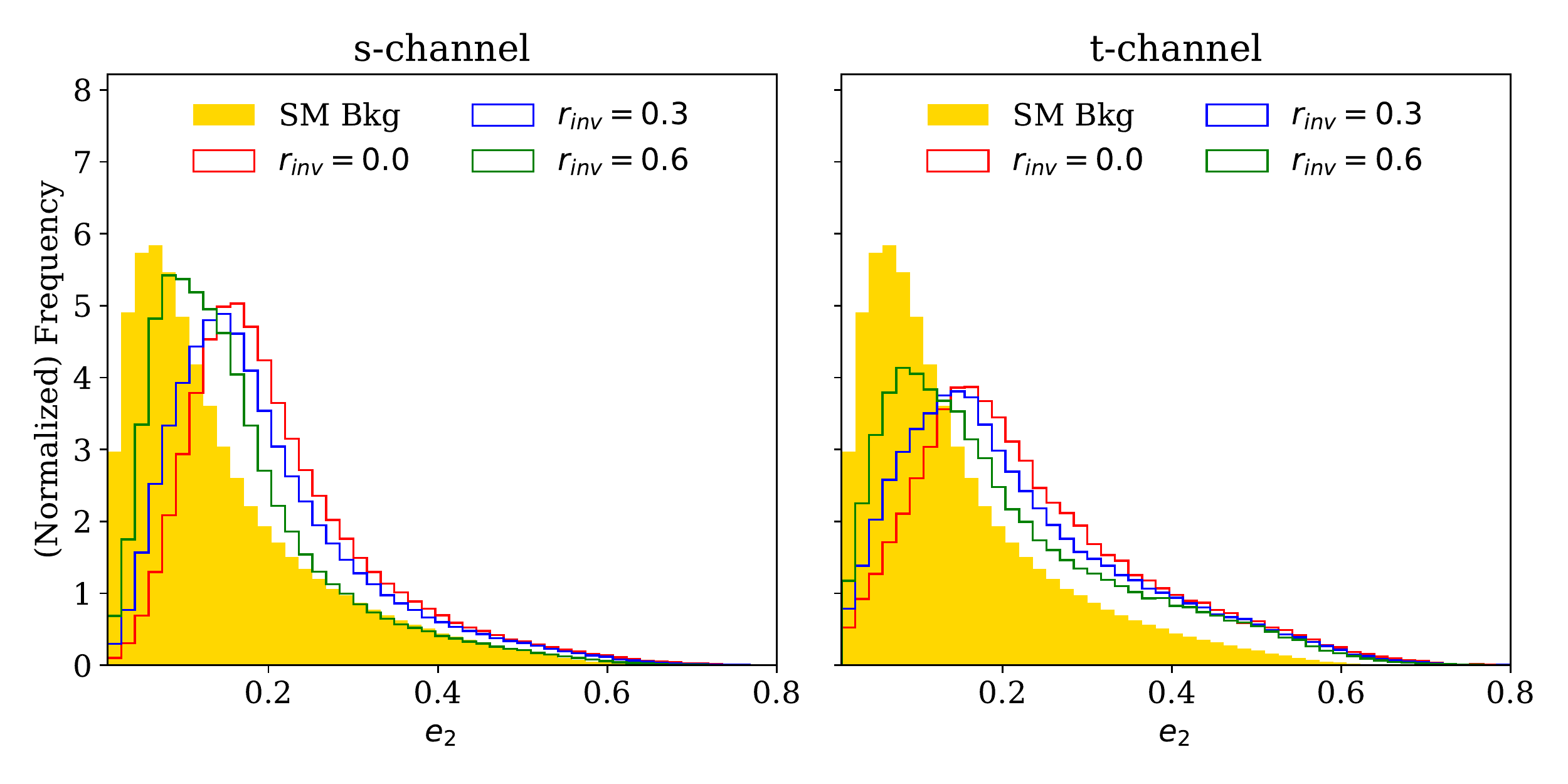}
	\includegraphics[width=0.48\textwidth]{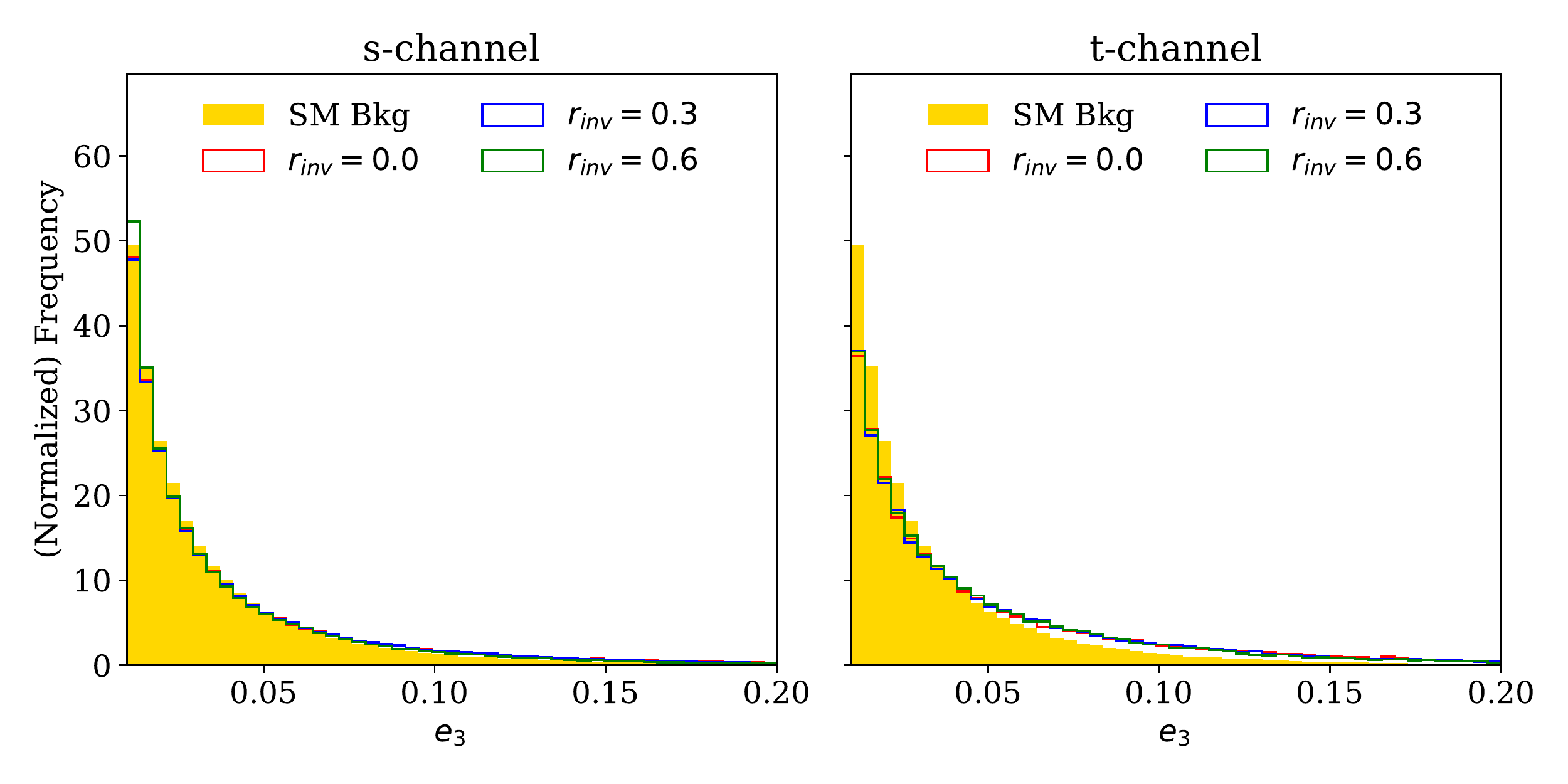}
	\caption{Distributions of jet energy correlation functions $\left(C_{2}^{\beta=1}, C_{2}^{\beta=2}, D_{2}^{\beta=1}\, \text{and}\, D_{2}^{\beta=2}\right)$ and pairs $e_2$ and $e_3$ shown for semi-visible jet (red, green, blue) and the standard model background (SM Bkg, yellow) for the six simulated scenarios, three choices of invisible fraction $\rinv$ for both the $s$-channel and $t$-channel processes.}
	\label{fig:ecf}
\end{figure*}

Energy Correlation Functions~\cite{Larkoski:2013uj} and their corresponding ratios are computed via the functions: $\ecf{1}, \ecf{2}$ and $\ecf{3}$,
\begin{align}
	\ecf{1}       & = \sum_i \pti{i}                                                                                     \\
	\ecf{2}^\beta & = \sum_{i<j} \pti{i}\, \pti{j} \left(\theta_{ij} \right)^{\beta}                                     \\
	\ecf{3}^\beta & = \sum_{i<j<k} \pti{i}\, \pti{j}\, \pti{k} \left(\theta_{ij} \theta_{ik} \theta_{jk} \right)^{\beta}
\end{align}
and the related ratios are given by,
\begin{align}
	e_2^\beta & = \frac{\ecf{2}^\beta}{\left(\ecf{1}\right)^2} \\
	e_3^\beta & = \frac{\ecf{3}^\beta}{\left(\ecf{1}\right)^3}
\end{align}
from these ratios, we then compute the energy correlation ratios $C_2$ and $D_2$
\begin{align}
	C_2 & = \frac{e_3}{\left(e_2\right)^2} \\
	D_2 & = \frac{e_3}{\left(e_2\right)^3}
\end{align}
Distributions for all Energy Correlation observables are shown in \Fig{fig:ecf}

\subsection{N-Subjettiness}
\begin{figure*}[ht]
	\centering
	\includegraphics[width=0.48\textwidth]{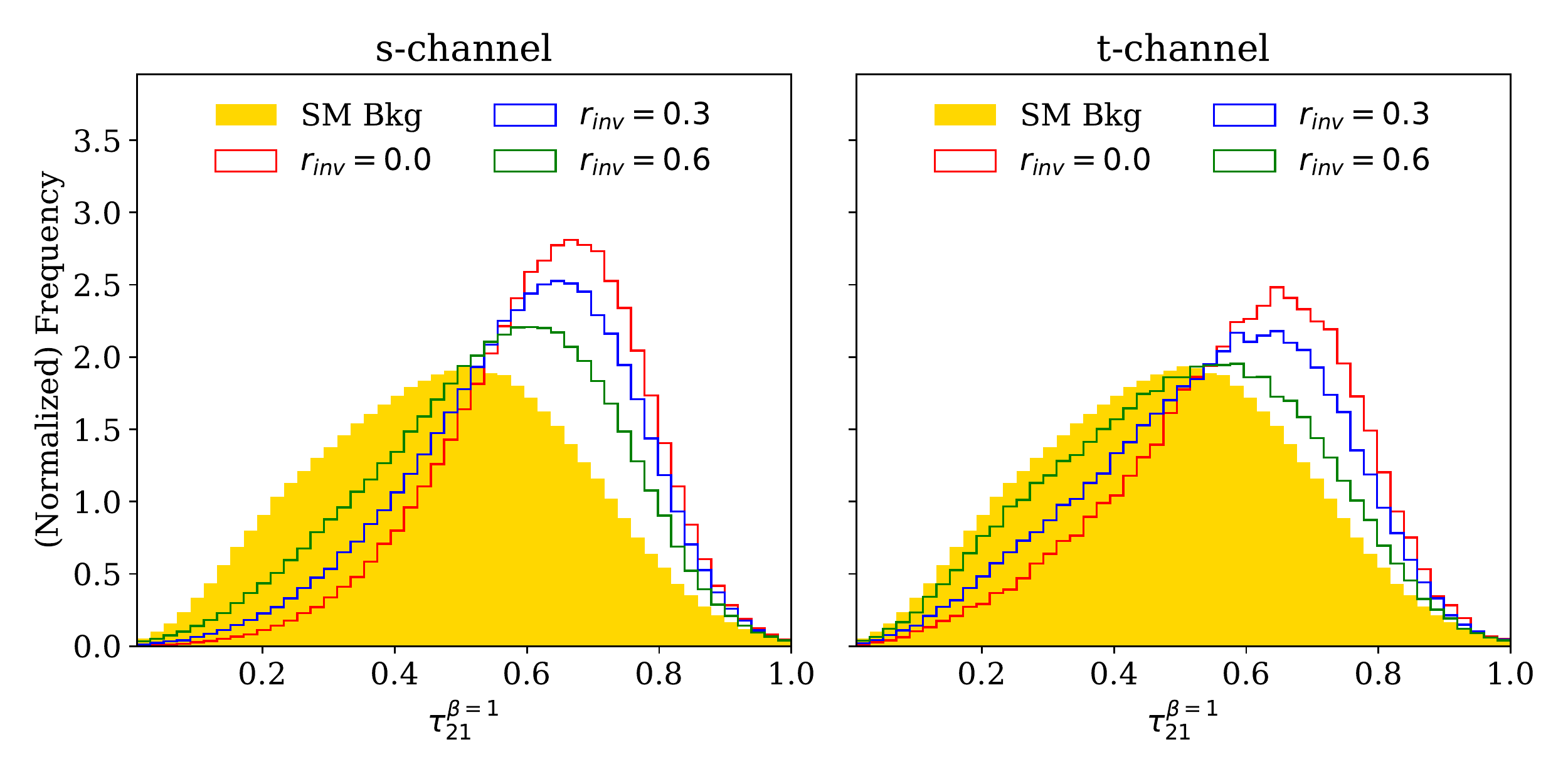}
	\includegraphics[width=0.48\textwidth]{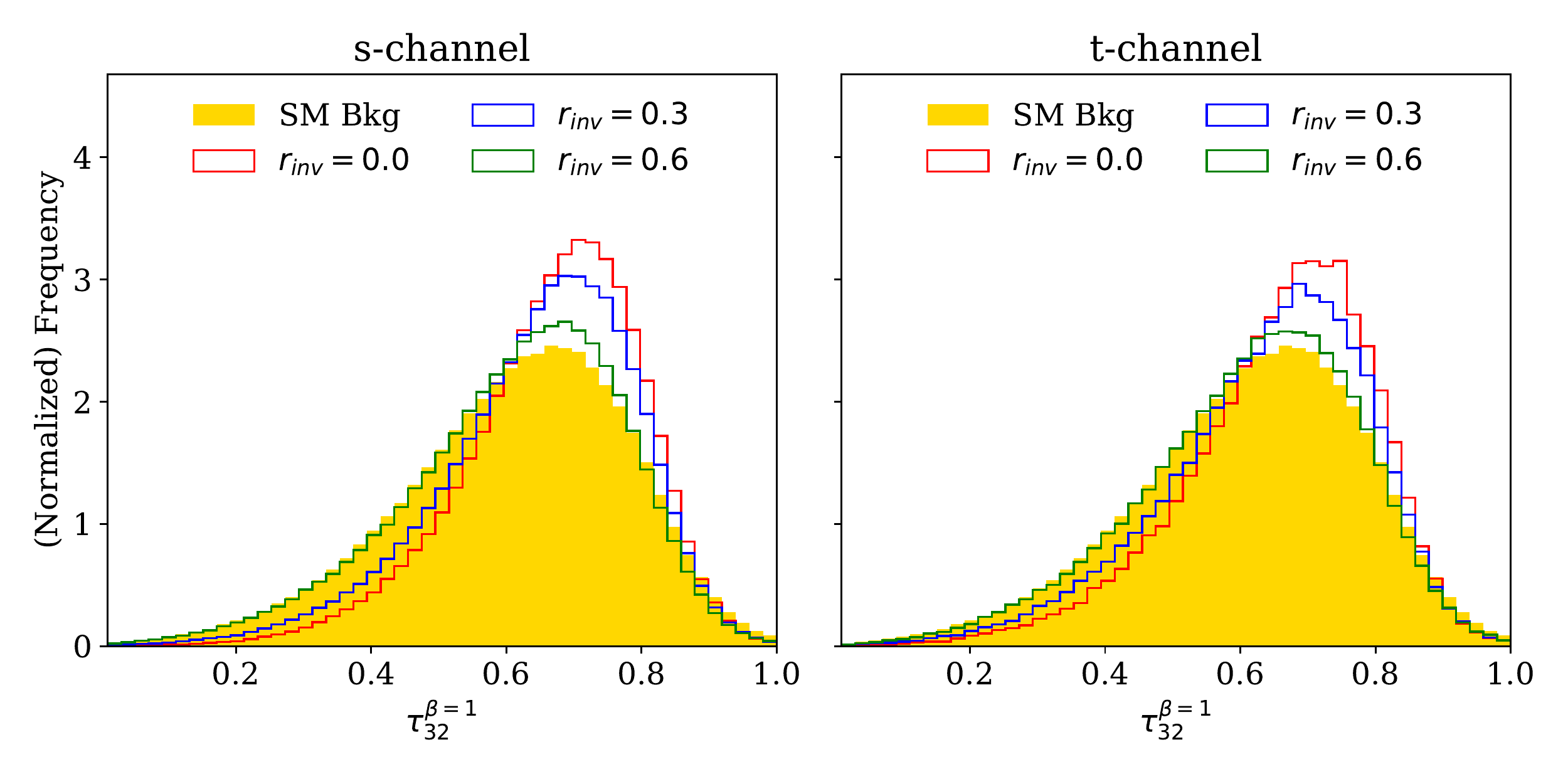}
	\caption{Distributions of jet N-subjettines $\left( \tau_{21}^{\beta=1}\, \text{and}\, \tau_{32}^{\beta=1}\right)$ shown for semi-visible jet (red, green, blue) and the standard model background (SM Bkg, yellow) for the six simulated scenarios, three choices of invisible fraction $\rinv$ for both the $s$-channel and $t$-channel processes.}
	\label{fig:nsubjet}
\end{figure*}

Given subjets isolated via clustering, for N candidate subjets, the N-subjettiness ($\tau_N$)~\cite{Thaler2011IdentifyingN-subjettinessb} is defined as,
\begin{equation}
	\tau_N = \frac{1}{d_0} \sum_k \pti{k} \textrm{min}\left( \Delta \theta_{1,k}, \Delta \theta_{2,k}, \ldots, \Delta \theta_{N,k} \right)
\end{equation}
where we define the normalization factor $d_0$ by,
\begin{equation}
	d_0 = \sum_k \pti{k}\, R_0
\end{equation}
where $R_0$ is the characteristic jet radius used during clustering. Finally, the N-subjettiness ratios used are defined by
\begin{align}
	\tau_{21}^{\beta=1} & = \frac{\tau_2}{\tau_1} \\
	\tau_{32}^{\beta=1} & = \frac{\tau_3}{\tau_2}
\end{align}
Distributions for both N-subjettiness observables are shown in \Fig{fig:nsubjet}

\subsection{Groomed Momentum Splitting Fraction}
The splitting fraction is described in terms of the Soft Drop grooming technique in \rref{Larkoski2014SoftDrop}. The feature is calculated using energyflow~\cite{Komiske:2017aww} with the Cambridge/Aachen algorithm using a jet radius of $R=1$ and Soft Drop parameters of $\beta=0$ and $z_{\text{cut}}=0.1$.

\begin{figure}[ht]
	\centering
	\includegraphics[width=0.48\textwidth]{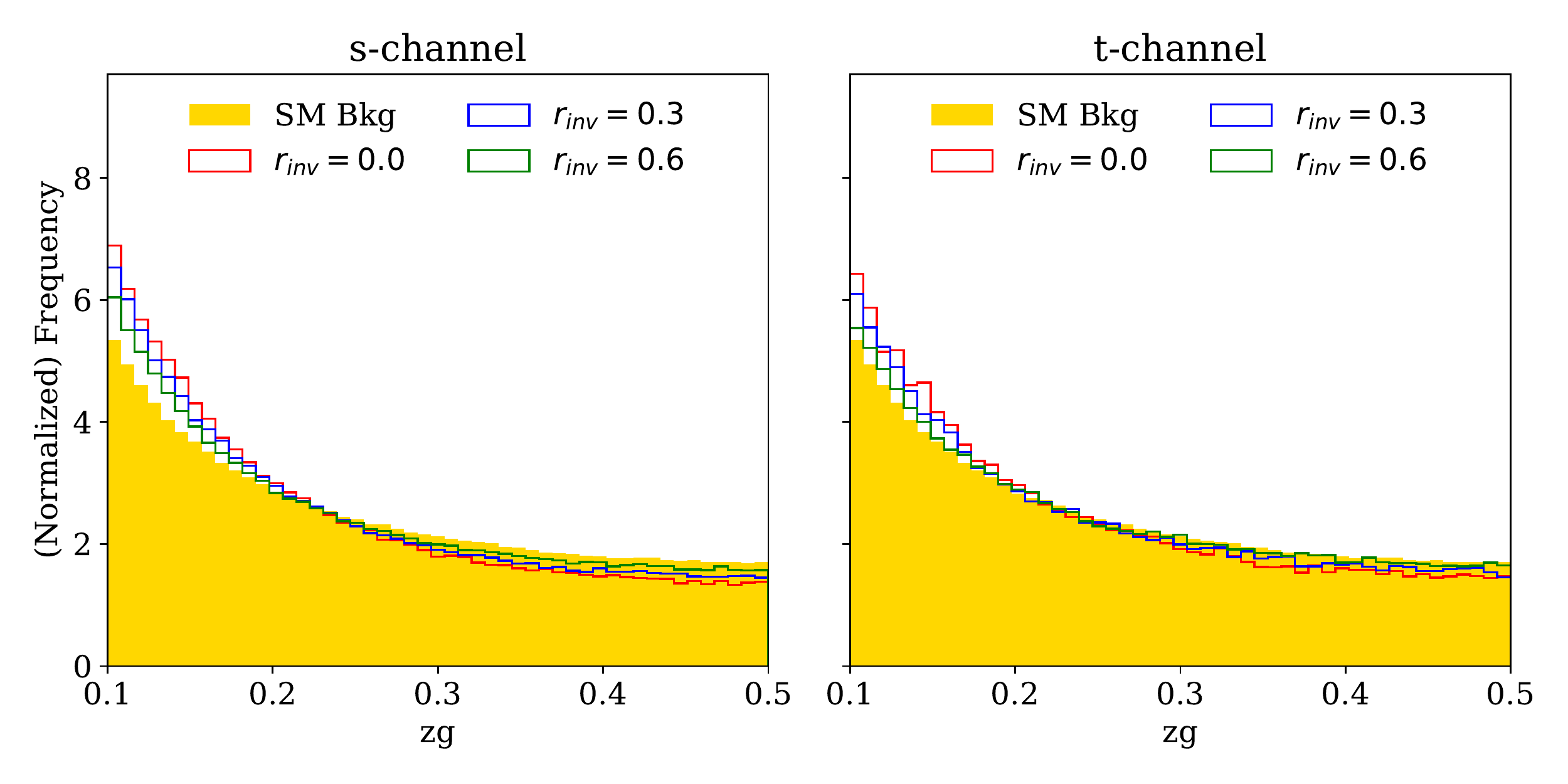}
	\caption{Distributions of jet splitting function $z_g$ shown for semi-visible jet (red, green, blue) and the standard model background (SM Bkg, yellow) for the six simulated scenarios, three choices of invisible fraction $\rinv$ for both the $s$-channel and $t$-channel processes.}
	\label{fig:zg}
\end{figure}

A distribution for $z_g$ is given in \Fig{fig:zg}

\section{ML Architectures}
\label{sec:ml_arch}

\subsection{Deep Neural Networks}
All deep neural networks were trained in Tensorflow~\cite{tensorflow2015} and Keras~\cite{chollet2015keras}. The networks were optimized with Adam~\cite{Adam2014} for up to 100 epochs with early stopping. For all networks, weights were initialized using orthogonal weights~\cite{OrthogonalWeights}. Hyperparameters were optimized using bayesian optimization with the Sherpa hyperparameter optimization library~\cite{hertel2020sherpa}.

\subsection{High-Level DNN}
\label{sec:hl_efp_networks}
All HL features are preprocessed with Scikit's Standard Scaler~\cite{Pedregosa2011Scikit-learn:Python} before training.

\subsubsection{Deep Neural Networks}
Hyperparameters and network design for all Dense networks trained on HL or EFP features are selected via Sherpa optimization using between two and eight fully connected hidden layers and a final layer with a sigmoidal logistic activation function to predict the probability of signal or background. 

\subsubsection{Particle-Flow Networks}
The Particle Flow Network (PFN) is trained using the energyflow package~\cite{Komiske:2018cqr}. Input features are taken from the trimmed jet constituents and preprocessed by centering the constituents in $(\eta-\varphi)$ space to the average $p_{\textrm{T}}$ and normalizing constituent values to 1. Both the EFN and PFN use this constituent information as inputs in the form of the 3-component \verb@hadronic measure@ measure option in EnergyFlow (i.e. $p_{\textrm{T}}$, $\eta$, $\phi$).

The PFN uses 3 dense layers in the per-particle frontend module and 3 dense layers in the backend module. Both frontend and backend layers use 300 hidden nodes per layer with a latent and filter dropout of 0.2. Each layer uses relu~\cite{Nair2010} activation and glorot normal initializer. The final output layer uses a sigmoidal logistic activation function to predict the probability of signal or background. The Adam optimizer is used and trained with a batch size of 128 and a fixed learning rate of 0.001.

\subsection{Boosted Learning Models}
HL features are, again, preprocessed with Scikit's Standard Scaler before training. Except where indicated, default settings are used.
\subsubsection{LightGBM}
All applications of LightGBM are trained using regression for binary log loss classification using Gradient Boosting Decision Trees. Performance is measured by the AUC metric for a maximum of 5000 boosting rounds and early stopping set to 100 rounds against AUC improvements.
\subsubsection{XGBoost}
All applications of XGBoost are trained using using the gradient tree booster and settings of: $\eta$=0.1, subsample=0.5, \verb@base_score@=0.1, $\gamma$=0.0, and \verb@max_depth@=6. Performance is measured by the AUC metric for a maximum of 5000 boosting rounds and early stopping set to 100 rounds against AUC improvements.

\subsection{Convolutional Networks on Jet Images}
The Convolutional Neural Networks used a jet image produced through EnergyFlow's \verb@pixelate@ function. Jet images were produced as a $32\times 32$ pixel matrix with an image width of 1.0. Resulting jet images were then normalized to a range of values between [0,1]. The network consisted of 3 hidden layers consisting of 300 nodes and used kernels of size $3 \times 3$ and strides of $1 \times 1$. Each layer uses relu~\cite{Nair2010} activation and glorot normal initializer. The final output layer uses a sigmoidal logistic activation function to predict the probability of signal or background. The Adam optimizer is used and trained with a batch size of 128 and a fixed learning rate of 0.001.
 
\bibliographystyle{apsrev4-1}
\bibliography{references}
\end{document}